# Direct Detection of Spin Polarization in Photoinduced Charge Transfer through a Chiral Bridge

Alberto Privitera[1,2], Emilio Macaluso[3,4], Alessandro Chiesa[3,4], Alessio Gabbani[5], Davide Faccio[6], Demetra Giuri[6], Matteo Briganti[2], Niccolò Giaconi[2,7], Fabio Santanni[2], Nabila Jarmouni[5], Lorenzo Poggini[8], Matteo Mannini[2], Mario Chiesa[1], Claudia Tomasini[6], Francesco Pineider[5], Enrico Salvadori[1*], Stefano Carretta[3,4*], Roberta Sessoli[2*]

[1.] *Department of Chemistry, University of Torino, Via Giuria 9, Torino, Italy*

[2.] *Department of Chemistry "U. Schiff" (DICUS), University of Florence,* & UdR INSTM Firenze, *Via della Lastruccia 3-13, Sesto Fiorentino, Italy*

[3.] *Department of Mathematical, Physical and Information Sciences, University of Parma, I-43124, Parma, Italy & UdR INSTM Parma*

[4.] *INFN–Sezione di Milano-Bicocca, gruppo collegato di Parma, I-43124 Parma, Italy.*

[5.] *Department of Chemistry and Industrial Chemistry, University of Pisa,* & UdR INSTM Pisa, *Via Moruzzi 13, Pisa, Italy*

[6.] *Department of Chemistry "Giacomo Ciamician", University of Bologna, Via Selmi 2, Bologna, Italy*

[7.] *Department of Industrial Engineering (DIEF), University of Florence, & UdR INSTM Firenze, Via Santa Marta 3, Firenze, Italy*

[8.] *CNR-ICCOM, Via Madonna del Piano 10, Sesto Fiorentino, Italy*

**ABSTRACT:** It is well assessed that the charge transport through a chiral potential barrier can result in spin-polarized charges. The possibility of driving this process through visible photons holds tremendous potential for several aspects of quantum information science, e.g., the optical control and readout of qubits. In this context, the direct observation of this phenomenon via spin-sensitive spectroscopies is of utmost importance to establish future guidelines to control photo-driven spin selectivity in chiral structures. Here, we provide direct proof that time-resolved electron paramagnetic resonance (EPR) can be used to detect long-lived spin polarization generated by photoinduced charge transfer through a chiral bridge. We propose a system comprising CdSe QDs, as a donor, and $C_{60}$, as an acceptor, covalently linked through a saturated oligopeptide helical bridge ($\chi$) with a rigid structure of ~ 10Å. Time-resolved EPR spectroscopy shows that the charge transfer in our system results in a $C_{60}$ radical anion, whose spin polarization maximum is observed at longer times with respect to that of the photogenerated $C_{60}$ triplet state. Notably, the theoretical modeling of the EPR spectra reveals that the observed features may be compatible with chirality-induced spin selectivity and identifies which parameters need optimization for unambiguous detection of the phenomenon. This work lays the basis for the optical generation and direct manipulation of spin polarization induced by chirality.

## Introduction

The second quantum revolution is unfolding now, exploiting the enormous advancements in our ability to detect and coherently manipulate single quantum objects.[1] In this context, the possibility of controlling the electron spin of molecular qubits[2-4] – or reading out the information encoded in their spin states – through the use of visible photons represents an attractive approach toward the realization of smaller, faster, and more energy-efficient QI and spintronic technologies.[5-7] An emerging possibility to address this challenge is the use of chirality.[8-10] Chiral structures have recently received significant attention thanks to their spin filtering behavior in the phenomenon described as Chirality-Induced Spin Selectivity (CISS).[11-12] This mechanism has been adopted to interpret a wide range of experimental results in which chiral systems impart significant spin selectivity in electron transport through chiral molecules, oligomers, and polymers.[8-15] Notably, the spin selectivity of the CISS effect is exceptionally high, even at room temperature.[16-17] In contrast to alternative methods previously used to achieve spin-to-light (or vice versa) interconversion, the CISS effect - which operates at the molecular scale – has the potential to reach the sensitivity for the readout of individual spins.[18] The latter appears even more interesting considering that the molecular flexibility achievable through chemical tunability allows controlling key features like the qubit-qubit interactions, crucial for implementing quantum gates.[19]

Despite the recent experimental and theoretical efforts aimed at rationalizing the origin and the potentialities of the CISS effect, the phenomenon is not fully understood yet. Most scientific results come from experiments performed on hybrid materials comprising chiral molecules supported on metallic substrates.[8, 11-12] Conversely, the implementation of the CISS effect in molecular charge transfer (CT) through the use of light has tentatively been tested only recently via indirect methods, e.g. probing competitive nonradiative and radiative relaxation processes as a function of an external magnetic field, light polarization, and molecular or helical handedness.[5] The challenge of directly detecting at the molecular level the non-Boltzmann spin populations that arise from photoinduced CISS originates from the lack of suitable donor-chiral bridge-acceptor (D-$\chi$-A) systems. The latter must simultaneously show good photoinduced CT efficiencies and efficient spin-filtering effect through the chiral bridge.[20] In this regard, systems comprising a semiconductor quantum dot (QD) as a donor and an organic molecule as an acceptor have recently demonstrated to provide a robust platform to transfer spin polarization from the photo-excited QD to the organic molecule, whose non-Boltzmann spin populations can be investigated in detail through the use of time-resolved electron paramagnetic resonance (trEPR) spectroscopy.[21-22] The development of a sound theoretical framework to rationalize the trEPR spectra of these systems is mandatory for investigating spin selectivity in the CT process through an effective chiral potential. A deeper understanding of spin selectivity would dramatically advance our capabilities to control and harness CISS at a fundamental scale.

Here, we directly probe the spin-polarized CT process through a chiral bridge via time-resolved EPR (trEPR) spectroscopy in a model system comprising CdSe QD-chiral bridge-fullerene (hereafter QD-χ-$C_{60}$). The studied chiral system is favorable for several reasons: (1) QDs are an effective reservoir of electrons that can be donated via CT towards $C_{60}$, (2) $C_{60}$ is a good electron acceptor, and (3) the used chiral bridge is a two-unit polypeptide belonging to the most investigated class of chiral linkers in CISS-based spintronic devices. The QD-χ-$C_{60}$ system is synthesized via a ligand exchange approach and characterized through the combination of optical spectroscopies and X-ray photoemission spectroscopy (XPS), which allows a detailed and quantitative elemental analysis and confirms the chemisorption of the $C_{60}$ chiral derivative. Time-resolved EPR spectroscopy shows that the photoinduced CT from the QD to the fullerene generates a spin-polarized state assigned to the fullerene radical anion. Detailed modeling of the EPR features indicates that they are compatible with the presence of CISS, albeit the unambiguous detection of CISS is hampered by the equilibration of spin population on the photo-excited QD. Though not yet conclusive, our investigation represents the first attempt to observe the photoinduced CISS effect directly. This provides a fundamental step to formulate clear and handy guidelines for designing future materials based on CISS and ultimately championing the burgeoning field of QI science.

## Results and discussion

**Chiral system engineering and synthesis.** The QD-χ-$C_{60}$ system investigated in this paper is shown in Figure 1a. We engineered our system to optimize the CT process, which is the main focus of our analysis. CdSe QDs represent a sound model system thanks to their tailorable light absorption (through their size) and excellent electron-donating properties.[23] Prior studies, based on different experimental methods, e.g. photoelectron spectroscopy in air (PESA) and theoretical calculations, have shown that the approximate valence and conduction band energies of CdSe QDs with a diameter of 5 nm are ~ -3.1 and -5.4 eV, respectively.[24] However, it should be noted that different techniques give different band energies, which are also strongly affected by the environment and the ligand nature. Thus, we represent these states as a distribution. The $C_{60}$ molecule is one of the best performing electron acceptors,[25] whose frontier energy levels, also distributed in energy, are approximately found at -4.0 and -6.2 eV.[26] The diagram in Figure 1b also reports approximate energetics and a possible path for the CT in our system. Specifically, after the photoexcitation of the QD, an electron transfer to the $C_{60}$ can occur due to the favorable energy levels alignment. If $C_{60}$ is photoexcited, a hole transfer will occur, resulting in the same final CT state (hole in the QD and electron in the $C_{60}$).

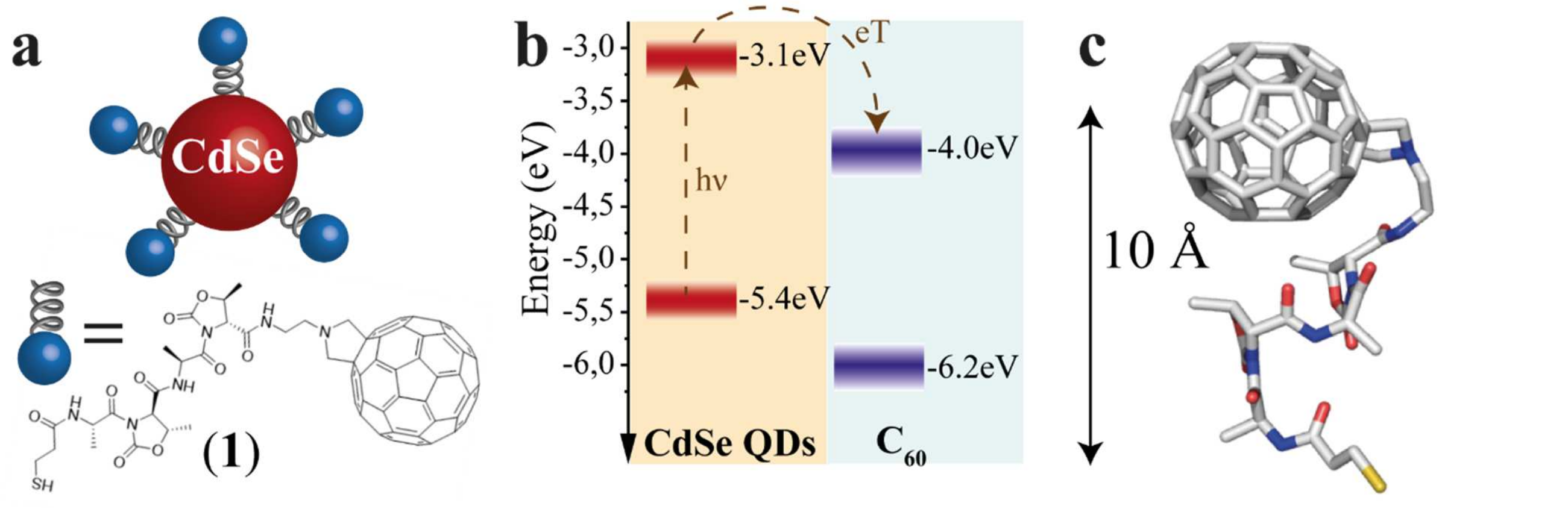

**Figure 1 – Engineering of the D-χ-A system: synthesis, energy levels, and DFT calculations of QD- χ-$C_{60}$.** (a) Schematic representation of the CdSe QD-χ-$C_{60}$ system. The QD acts as an electron donor and $C_{60}$ as an electron acceptor. The chiral bridge is characterized by a two-units peptidic chain, which offers an effective chiral potential through which the CT process occurs. (b) Energy level diagram of CdSe QD and $C_{60}$. The dotted arrows represent the CT process after light absorption by the CdSe QD. Conversely, a hole transfer process can occur if the $C_{60}$ absorbs the light. Both processes generate a hole localized on the CdSe QDs and a radical anion localized on the $C_{60}$. (c) DFT calculations of the ligand **1**. The calculations highlight the presence of a rigid structure with a distance from the S atom to the $C_{60}$ of about 10Å.

The selected linker, reported in Figure 1a, is composed of two units comprising L-Alanine and the pseudo amino acid (4*S*,5*R*)-4-methyl-5-carboxyl-oxazolidin-2-one (D-Oxd). The grafting thiolate group is on the N termination of the peptide chain, while $NH_2(CH_2)_2$-fulleropyrrolidine is bound to the C-termination of the polypeptide to give SH-$(CH_2)_2$-(CO)-(L-Ala-D-Oxd)$_2$-NH-$(CH_2)_2$-fulleropyrrolidine, **1**. This choice aims to find an acceptable compromise between CT efficiency and effective chiral potential for the CT process. In a system where CT occurs in all directions, CISS can still be observed, as suggested by our previous work,[18] though it is independent from the chosen enantiomer. Thus, we focused our analysis only on one of the two possible ligand enantiomers. Since the X-Ray structure of the adduct **1** is not accessible, to clarify the geometrical conformation of the chiral bridge after $C_{60}$ grafting, we performed density functional theory (DFT) geometry optimization of the isolated structure of **1** (see Computational Details in SI). In Figure 1c, the final optimized structure is reported. Minor deviations from the available X-ray structure of the S-protected oligopeptide lacking the fullerene unit [Tri-S-$(CH_2)_2$-(CO)-(L-Ala-D-Oxd)$_2$-OBn][27] have been computed (see figure S1). The small root mean square deviation (RMSD) of 0.516 Å between the optimized structure of the oligopeptide moiety in **1** and the crystal structure of the free oligopeptide suggests that the geometry of the pristine chiral bridge is preserved even after the bonding to the fulleropyrrolidine. This can be rationalized considering that the chiral bridge is relatively rigid despite its shortness and the absence of intermolecular interactions active in the crystal. For this reason, we can conclude that the intramolecular hydrogen bonds between the two peptide units of the linker are enough to stabilize the adduct **1**, even in the isolated structure. As a result, we can infer that the structure of **1** is reasonably preserved also upon grafting on the QD. In addition, the insertion process of individual **1** ligands into a dense layer of capping molecules should prevent the folding of $C_{60}$ on the QD.[28-30] This is further corroborated by the observation of a trEPR signal (vide infra), suggesting slow charge recombination and, therefore, a significant distance between the QD and $C_{60}$.[21]

After thoughtfully selecting the chiral linker, we turn to the synthesis and experimental characterization of the proposed QD-χ-$C_{60}$ system. CdSe QDs were prepared via the hot injection method reported by Dai *et al.* with minor modifications.[31] The average diameter of the QDs is ~ 5 nm, as shown by TEM analysis. Further details on the synthesis and morphological characterization are reported in SI.

The fullerene-functionalized polypeptide **1** was prepared through liquid phase synthesis. The D-Oxd-OBn group was prepared from threonine and coupled with Boc-L-Ala.[32] Fullerene-$C_{60}$ was derivatized according to the Prato reaction to afford *N*-2-aminoethyl-fulleropyrrolidine.[33] Then, Boc-(L-Ala-D-Oxd)$_2$-OBn was derivatized by replacing the OBn protecting group with the previously obtained *N*-2-aminoethyl-fulleropyrrolidine and the *N*-Boc protecting group with 3-(tritylthio)propanoic acid by standard coupling reactions. Finally, the S-protecting trityl group was removed,[34] and compound **1** was obtained. For further details, see SI.

The ligand-exchange reaction of the as-synthesized CdSe QDs capped with trioctylphosphine oxide (TOPO) was favored by the strong affinity of thiols to the CdSe surface, according to HSAB theory for an X-type ligand.[35] We performed ligand exchange by adding an excess of **1** (10 times higher than the estimated quantity for maximum coverage of the QD surface) to a solution of QDs in chloroform and keeping the mixture overnight under mechanical stirring. After the exchange reaction, the newly-formed QDs (QD-χ-$C_{60}$ system) precipitate, as the presence of $C_{60}$ in the ligand drastically reduces their colloidal stability. The change in solubility is the first evidence of the successful exchange of the native ligands with **1**. The QD-χ-$C_{60}$ system was subsequently re-dispersed in a solution of 1,2,4-trichlorobenzene and purified (details in SI). For comparative studies, two additional molecular systems were synthesized: i) SH-$(CH_2)_2$-(CO)-(L-Ala-D-Oxd)$_2$-L-Val-OMe (**2**) to obtain a similar functionalization of the QDs without the $C_{60}$ acceptor; ii) fulleropyrrolidine-$(CH_2)_2$-NH-COO*t*Bu lacking the thiol grafting group (**3**). For the molecular structures, see Schemes S1 and S2 in the SI.

To confirm the success of the ligand-exchange reaction, we started by carrying out optical measurements. In particular, we used UV-vis absorption, steady-state PL, and transient PL to investigate the photophysics of our model system as an initial platform on which to build the study of the spin dynamics mediated by light. In Figure 2a, we compare the UV-vis and PL spectra of the pristine CdSe QDs and the CdSe QD-χ-$C_{60}$ system. The UV-vis absorption spectrum of the CdSe QD-χ-$C_{60}$ system can be rationalized as the sum of two main contributions. The first originates from the CdSe QDs and shows a clear excitonic peak at around 600 nm, in analogy with the pristine CdSe QDs. The second is a broad absorption tail extending up to 700 nm typical of $C_{60}$ derivatives[36] and strongly resembles the UV-vis absorption spectrum of the free **1** ligand shown in Figure S16. The PL spectra provide a first fingerprint of the success of the ligand exchange reaction. Specifically, the inset of Figure 2a shows that the PL intensity is strongly quenched in the CdSe QD-χ-$C_{60}$ system with respect to the pristine CdSe QDs. This quenching is mainly ascribed to the thiol-mediated hole trapping process becoming dominant over radiative recombination.[37] A comparable PL quenching was obtained for CdSe QDs functionalized with **2**, where the $C_{60}$ molecule is absent, thus confirming the predominant role of thiols in the PL quenching mechanism (see Figure S17).

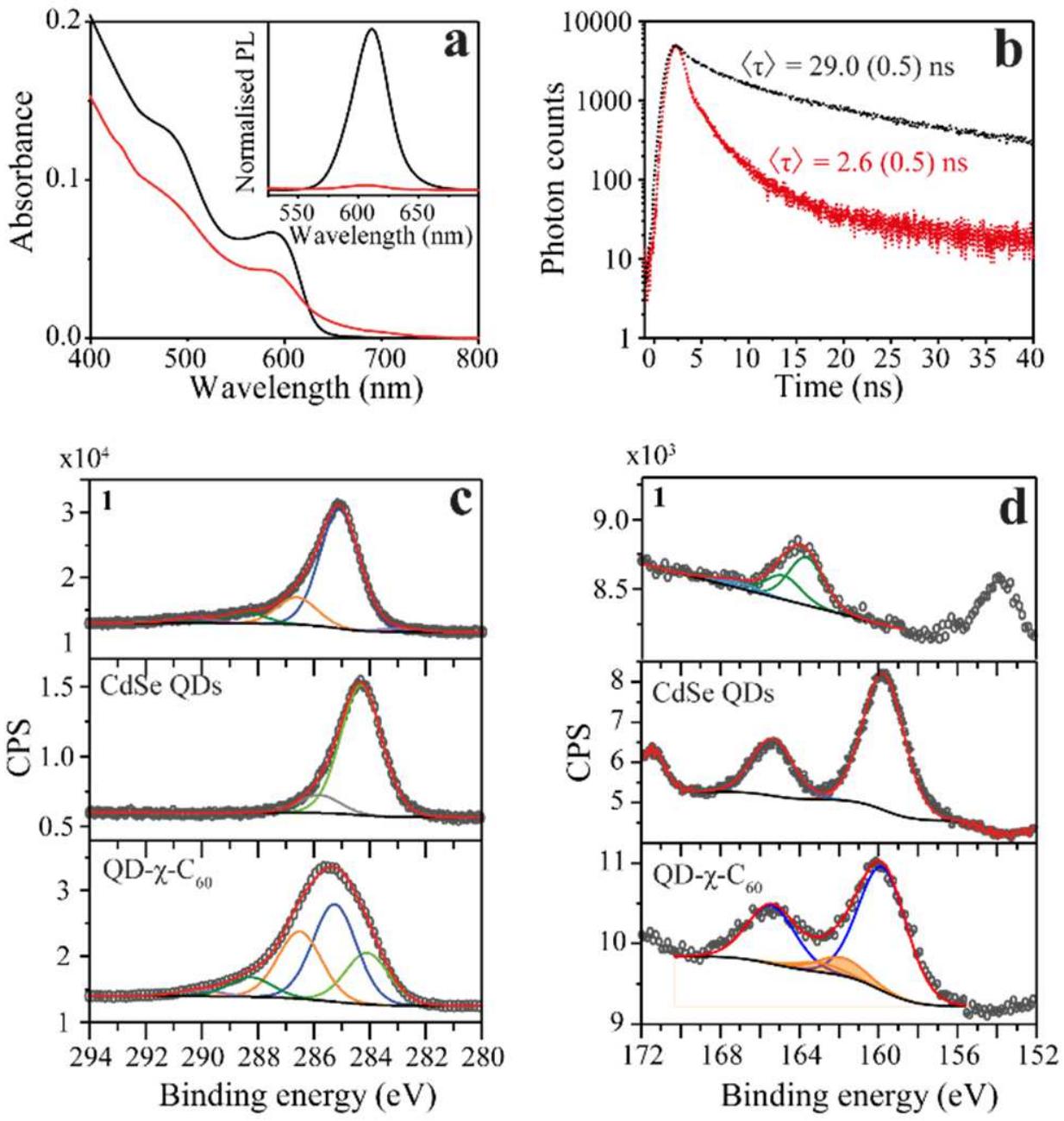

**Figure 2 – Optical and XPS characterization of the QD-χ-$C_{60}$ system.** (a) UV−vis absorption and normalized photoluminescence spectra excited at $\lambda_{ex}$=400 nm (inset) of pristine QDs (black line) and CdSe QD-χ-$C_{60}$ (red line) in 1,2,4-trichlorobenzene solution. Notably, the PL of the CdSe QD-χ-$C_{60}$ system is drastically quenched as a result of the thiol binding and probably of the CT process. (b) Photoluminescence decay curves recorded at λ=610nm with $\lambda_{ex}$=370 nm. From the fitting, an average decay time $\langle\tau\rangle$ of 29.0 (0.5) ns and 2.6 (0.5) ns were obtained, respectively, for the pristine and the chiral system. (c) C1*s* and (d) S2*p*/Se3*p* photoemission lines for ligand **1**, the CdSe QDs, and the QD-χ-$C_{60}$ system, as well as the single chemically shifted components from fit deconvolution.

We further probed the role of **1** ligand on the QD exciton decay by performing time-resolved photoluminescence (trPL) decay measurements on the QDs before and after the ligand exchange. The luminescence decay curves recorded at the emission maximum (Figure 2b) show a multiexponential decay that we modeled using triexponential decay kinetics, as shown in equation (1):

$$PL(t) = a_1 \exp\left(-\frac{t}{\tau_1}\right) + a_2 \exp\left(-\frac{t}{\tau_2}\right) + a_3 \exp\left(-\frac{t}{\tau_3}\right) \quad (1)$$

The fitting results are reported in Table S1 in the Supporting Information. The explanation for the origin of the multiexponential PL decay has been thoroughly discussed in the literature.[38] The main reason is the presence of QD surface defects that give rise to trap states lying within the bandgap and affecting the emission dynamics.[39-41] As a result, the photogenerated hole-electron pair exciton can follow different decay paths.[39] The most direct one consists of a rapid relaxation of the hole and the electron to the bottom of the valence and conduction bands, respectively, followed by the radiative relaxation to the ground state. This process contributes to the fastest lifetime decay. However, the hole (or the electron) can be localized in shallow trap states. These trapped charges can either repopulate the valence (or conduction) band or thermalize into deeper trap states. The former case contributes to the longer PL lifetimes, while the latter contributes to nonradiative mechanisms. The combination of all these processes and differences between the individual QDs give rise to multiexponential emission dynamics that occur over a nanosecond time scale, in agreement with literature values for similar systems.[38] In addition, the possibility of charge transfer from the CdSe QD to the $C_{60}$ fullerene in the QD-χ-$C_{60}$ system can also occur, as further observed below. To compare the PL lifetimes of the QDs before and after the ligand exchange, we calculated their average PL lifetime $\langle\tau\rangle$ using equation 2:[38]

$$\langle\tau\rangle = \sum_i a_i\tau_i^2 / \sum_i a_i\tau_i \qquad (2)$$

The results are $\langle\tau\rangle$ = 29.0 (0.5) ns to 2.6 (0.5) ns for the pristine CdSe QDs and the CdSe QD-χ-$C_{60}$ system, respectively. As expected from steady-state PL, the significant decrease in the PL intensity after ligand exchange confirms the fast nonradiative decay process induced by the thiol capping molecules, thereby confirming the success of the exchange interaction.[37]

We gathered further information on the surface of the functionalized QDs with XPS experiments. We analyzed the C1*s* and S2*p*/Se3*p* regions of the pristine CdSe QD and the QD-χ-$C_{60}$ system since they are the most relevant regions to investigate the chemical functionalization of the CdSe QDs. The analysis of Se3*d* and Cd3*d* regions is reported in SI (figures S18-22), together with the C1*s* and S2*p* XPS of the bulk phase of the individual building blocks (**1**, **2**, and $C_{60}$). The spectra of the C1*s* region are reported in Figure 3c and S18. In the spectrum of pristine CdSe QDs, the main component at 284.3 eV is attributed to aliphatic carbon atoms of the TOPO ligand, plus a minor component at 285.8 eV attributable to adventitious carbon.[42] Conversely, the *C*1*s* region acquired on the QD-χ-$C_{60}$ system features the components of both **1** and TOPO ligands (reported in figures S18), thereby demonstrating the coexistence of both ligands on the surface of CdSe QDs after the exchange reaction. However, the components belonging to TOPO decrease after the ligand exchange. Further confirmation of the presence of both ligands on the QDs surface is given by the P2*p* XPS signal (132.4 eV) detectable on samples before and after the exchange reaction (see Figure S20).[43]

Crucial insight regarding the assembly of molecules on the surface of the CdSe QDs can be deduced by the analysis of the S2*p*/Se3*p* region (Figures 2d and S19). In the pristine CdSe QDs sample, the Se3*p* signal (159.7 eV) and its relative spin-orbit contribution (+5.7 eV) are clearly visible.[44] In the QD-χ-$C_{60}$ system, we observe a change in the lineshape of the Se3*p* signal due to the overlap with additional components at 161.8 eV (highlighted in orange) attributable to sulfur atoms bound to the surface of the QDs.[45] Furthermore, the spectra do not feature signals at 163.5 eV and ca. 167 eV, which are characteristic of S-H and S-$O_n$ groups, thus excluding the presence of both physisorbed and oxidized species of **1**.

**Spin-polarized photoinduced charge transfer.** With the synthesis of the CdSe QD-χ-$C_{60}$ system confirmed, we turned to trEPR spectroscopy to investigate the photoinduced CT and its spin dynamics.[46-47] trEPR is sensitive to the presence of spin-polarized states, which show signals in enhanced absorption (*A*) and/or emission (*E*),[48] and, under favorable circumstances, can reach a time resolution as low as tens of nanoseconds. In trEPR spectroscopy, the detected signals result from non-Boltzmann population of the spin sublevels following the CT process.[20] The generation and time evolution of spin polarization is very informative about the spin dynamics of paramagnetic states.[49-50] We performed trEPR measurements at 40 K on our model system CdSe QD-χ-$C_{60}$ (7.8 μM in 1,2,4-trichlorobenzene, blue line). For comparison, we also investigated a solution containing both CdSe QD functionalized with **2** (CdSe QD-χ) (7.8 μM) and 1 mM of **3**. The concentration of QDs was chosen to have an optical density of 1 in the 0.33 cm EPR quartz tube. In Figure 3a, we show the trEPR spectra taken at 1 μs after excitation at 450 nm. In both spectra, we observe a broad signal between 335-358 mT that we assign to the $C_{60}$ triplet state, based on best-fit spectral simulations (Figure S23). From the simulation, a spin-polarization pattern typical of triplet states formed via intersystem crossing (ISC) promoted by spin-orbit coupling (SOC) is observed. This signal results from photogenerated singlet states in $C_{60}$ that do not undergo electron transfer. The $C_{60}$ spin populations in the two samples are slightly different, most likely due to a different environment of the $C_{60}$ molecule: linked to the CdSe QDs in the target system or dispersed in the frozen solution in the control experiment.

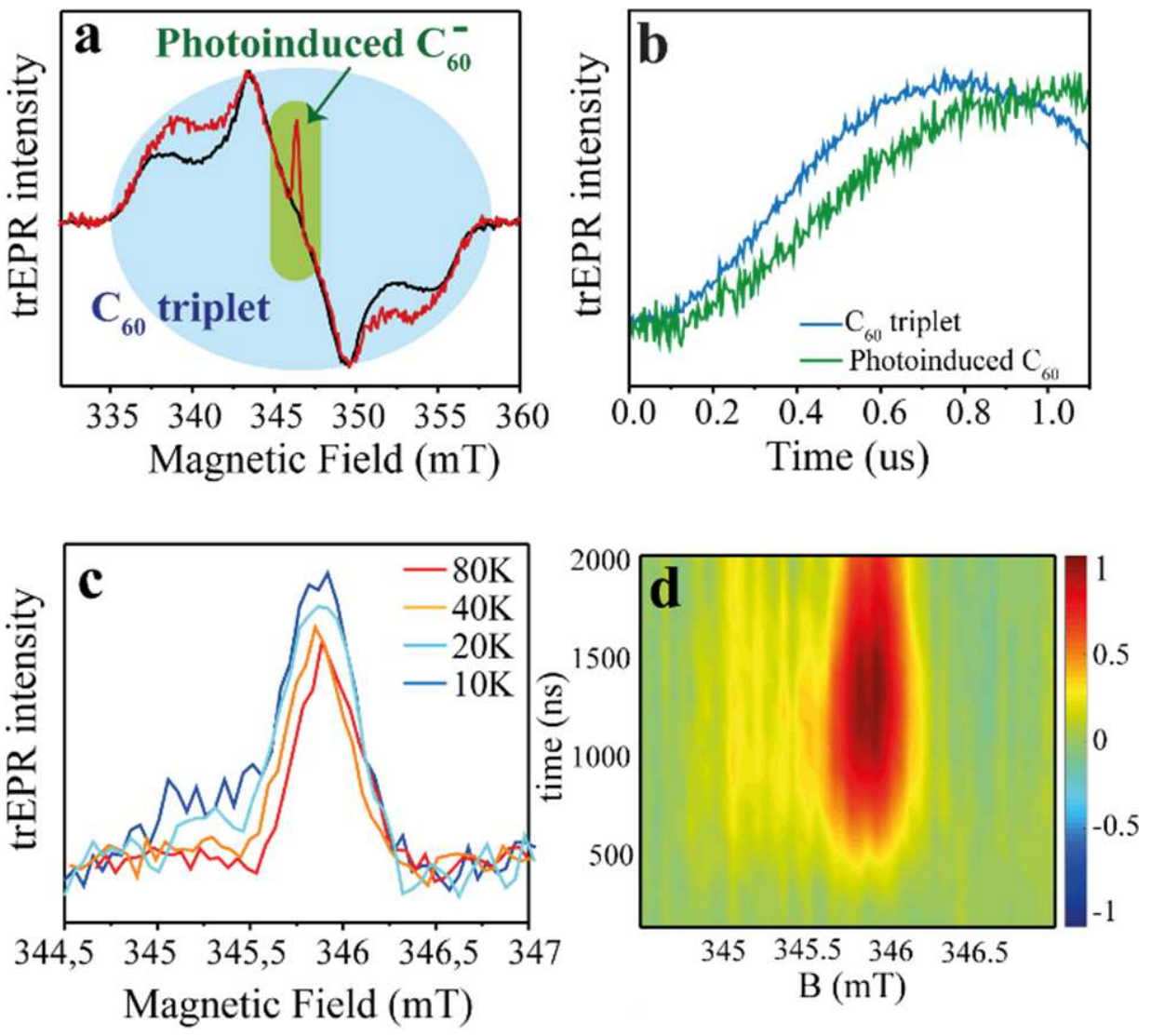

**Figure 3 - Time-resolved EPR spectra of the QD-χ-$C_{60}$ (model) and QD-χ (test) systems.** (1) Experimental trEPR spectra of QD-χ-$C_{60}$ (red line) and QD-χ + $C_{60}$ (black line) taken at 1 μs after 450 nm laser pulse (7ns, 2mJ) acquired at 40 K. Both spectra show a broad signal between ~ 335-358 mT, which is assigned to the $C_{60}$ triplet, but only the spectrum of QD-χ-$C_{60}$ displays a narrow signal in enhanced absorption centered at ~ 346 mT, which is attributed to the photogenerated $C_{60}$ radical anion. (b) Normalized trEPR transients of QD-χ-$C_{60}$ excited at 450 nm (40 K). The transients are integrated on a magnetic field window of 0.5 mT and are centered at 346 and 343 mT for the $C_{60}$ triplet and the photoinduced $C_{60}$ radical anion, respectively. The transients show a slower generation of the photoinduced $C_{60}$ radical anion with respect to the $C_{60}$ triplet. (c) Temperature-dependent trEPR spectra of QD-χ-$C_{60}$ taken at 1 μs after 450 nm laser pulse (7ns, 2mJ). The spectra show a mostly enhanced absorptive signal associated with the $C_{60}$ radical anion. (d) 2D experimental trEPR contour plot of QD-χ-$C_{60}$ of the charge transfer signal acquired after 450 nm laser pulse (7ns, 2mJ) at T = 10K. Color legend: red=enhanced absorption, blue=emission, green=baseline.

More interestingly, only in the spectrum of QD-χ-$C_{60}$ we observe an additional intense and spectrally narrow feature in enhanced absorption centered at ~346 mT (g-value ≈ 2.00), which we attribute to the $C_{60}$ radical anion.[22, 51] We propose that this absorptive signal results from the photoinduced electron transfer from the QDs to $C_{60}$. Specifically, after the absorption of 450 nm light by the CdSe QDs, an exciton is generated, which undergoes a CT process thanks to the favorable energy alignment (see Figure 1b).[22] As a result, a hole localized on the CdSe QD and a radical anion localized on the $C_{60}$ are formed. In agreement with literature reports,[21] the signal of the counterpart hole on the CdSe QDs is not visible in our spectra due to the fast spin relaxation induced by the large spin-orbit coupling of heavy Cd atoms. Lack of the trEPR signature of the hole was similarly reported by Olshansky et al..[21] As for the radical anion, the enhanced absorption polarization reminds of that recently observed in similar systems.[21, 52] However, a significant difference between our results and previous literature is the time evolution of spin polarization in the first μs after the laser pulse, reported in Figure 3b. In fact, in our case, the spin polarization of the $C_{60}$ signal rises even more slowly than the polarization of the $C_{60}$ ISC triplet and shows a maximum at ~1 μs, as further discussed in the theoretical modeling. In addition, most trEPR investigations of the CT process in similar QD/organic molecule systems involve D and A species that are nearby or connected by conjugated bonds,[21-22] while, in our case, the CdSe QD and the $C_{60}$ are covalently attached through a ~10Å long saturated bridge which in principle hinders the electron transfer process. Notably, our trEPR observation appears even more interesting considering the presence of a chiral bridge which has been suggested to favor the efficiency of the CT process.[5, 12]

To achieve better insight into the CT process, we investigated the signal of the spin-polarized $C_{60}$ anion by performing experiments at different temperatures (10 - 80 K), as reported in Figure 3c. The purely absorptive $C_{60}$ anion signal evolves at increasing temperatures, showing a weak emissive feature at lower fields. Polarization patterns of similar systems were rationalized in literature by considering two main contributions: (1) a main absorptive feature that originates from the triplet excited state of the photoexcited QD from which the CT process starts, and (2) a minor absorption/emission (*AE)* contribution showing up at lower temperatures (~10K) which originates from the spin-correlated radical pair (SCRP) mechanism.[21] However, distinctive spin polarization patterns and trends are observed in our case. In fact, as evidenced in Figure 3c, the emissive signal appears at low fields (*EA* pattern) and high temperatures (>40K). Since the origin of spin polarization in our and similar systems is still little understood, it is of paramount importance to theoretically understand the time evolution of the spin polarization of photoinduced CT states. This study appears even more fundamental in view of a rationalization of the role of the chiral linker in the spin-selectivity of the photoinduced CT process.

**Simulation of trEPR spectra.** In order to gain a deeper understanding of the CT process, we simulated trEPR spectra as a function of both time and static magnetic field.[53-57] Theoretical modeling focuses solely on the magnetic field region relevant for the charge transfer signal since the signal associated with the $C_{60}$ triplet (see Figure 3b) is due to a spin-polarized triplet state originating from SOC-promoted ISC independent of CISS. In our simulation, the initial state of the radical pair is described by the density matrix $\rho(0)$, written in the four-level basis composed by singlet and triplet spin states for the hole-electron pair. The time evolution of $\rho(0)$ is computed using the stochastic Liouville equation, considering both coherent and incoherent contributions.

$$\frac{\partial \rho}{\partial t} = -i[H,\rho] - \boldsymbol{L}\rho \qquad (3)$$

Coherent evolution is determined by the Hamiltonian $H$, which in the high-field approximation can be written in the rotating frame as follows:

$$H = \mu_B \boldsymbol{B}_0 \cdot (g_D \boldsymbol{S}_D + g_A \boldsymbol{S}_A) + \boldsymbol{S}_D \cdot \boldsymbol{D}(\Omega) \cdot \boldsymbol{S}_A - \hbar\omega_0 (g_D S_z^D + g_A S_z^A) + \mu_B B_1 (g_D S_x^D + g_A S_x^A\ ) \quad (4)$$

Where $g_D$ and $g_A$ are the isotropic g-factors of the donor QD and the $C_{60}$ acceptor radical species, $\mathbf{B}_0 = (0,0,B_0)$ is the static magnetic field, $\mathbf{D}$ is the spin-spin interaction tensor (including in principle both isotropic and dipole-dipole contributions), $\omega_0/2\pi =$ 9.69 GHz is the microwave frequency and $B_1 = 0.02$ mT is the microwave field strength. The spin-spin coupling is much weaker than the Zeeman energy in the examined system, which fully justifies the high-field approximation. Here we have modeled for simplicity the hole on the QD as an isotropic spin ½. This framework can be further extended to a more complex spin structure of the hole; however, this would require an extensive characterization of the QD, which is beyond the scope of the present work.

Incoherent evolution is accounted for by the super-operator **L**, which includes the effects of charge recombination from the singlet spin state (with rate $k_{CR}$), spin relaxation, and dephasing. Since the two $g$-factors are very different (see below), the eigenstates of the system are practically factorized. In addition, we expect a much shorter relaxation time for the electron spin on the QD. Hence, it is reasonable to assume two different relaxation times ($T_1^D$ and $T_1^A$) for the two electrons of the radical pair. Finally, the dephasing time $T_2$ (assumed to be the same for all transitions) induces a Lorentzian broadening of the EPR peaks.

By using the $\rho(\Omega, t)$ obtained for each time and magnetic field value, we compute the spherical average to obtain the trEPR spectrum:

$$EPR(t) = \int Tr\{(g_D \hat{S}_y^D + g_A \hat{S}_y^A)\hat{\rho}(\Omega,t)\} d\Omega \qquad (5)$$

The result of Eq. (5) is then convoluted with the exponential response function of the spectrometer, determined by its Q-factor of 6800, giving rise to a response time $t_R = 2Q/\omega_0 = 225$ ns.[58] Simulated spectra are plotted as a function of both time and magnetic field in Figure 4. For a better comparison, in Figure 3d we report the 2D experimental trEPR acquired at 10K as a contour plot similar to the computed ones.

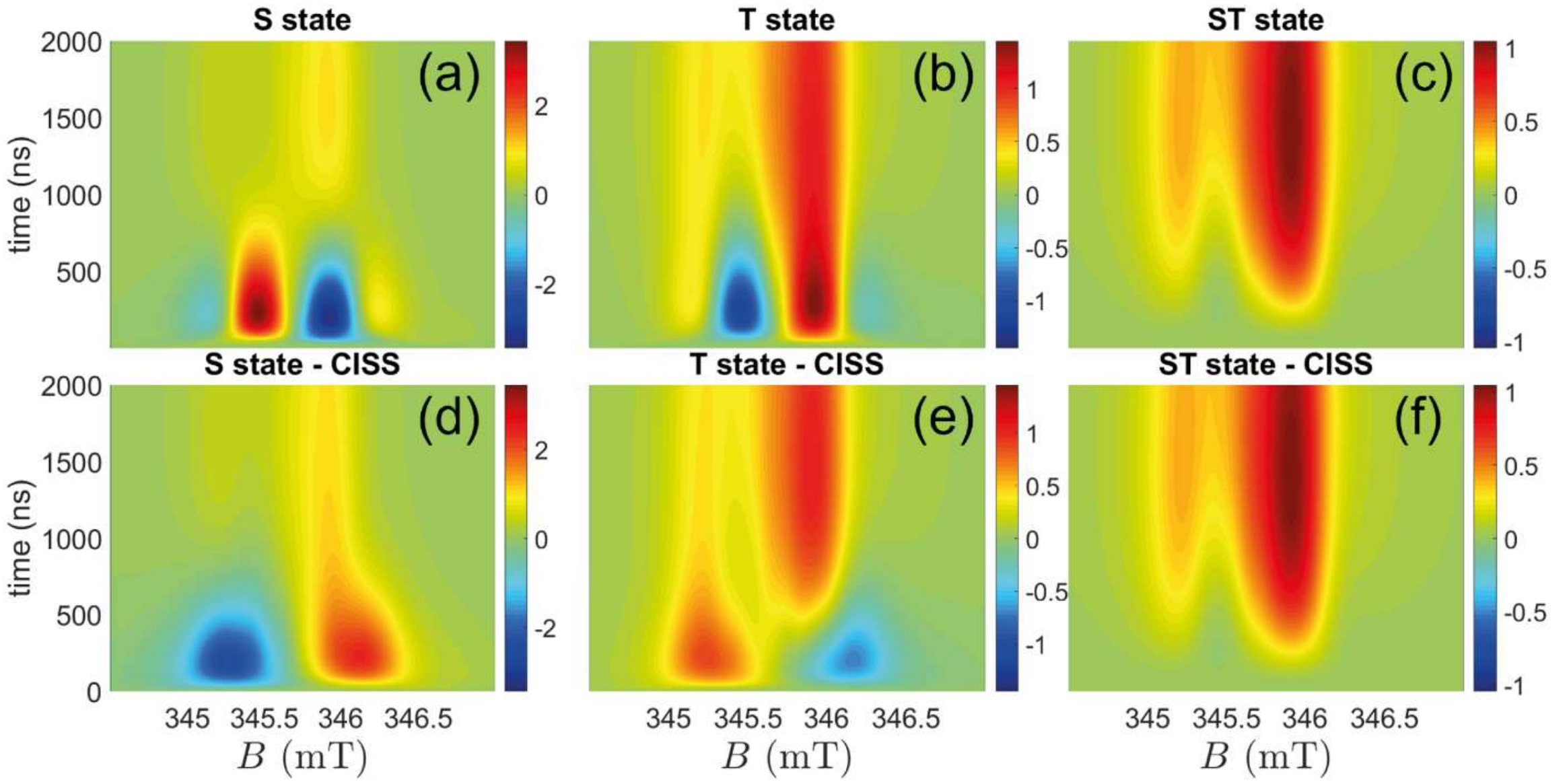

**Figure 4 – Simulated time-resolved EPR spectra as a function of both time and field**. Panels (a,b,c): spectra calculated for the S, T, and ST initial states. Panels (d,e,f): corresponding spectra calculated for the S, T, and ST initial states, filtered by the CISS effect. Simulations are performed by considering a temperature of *T*=10 K and using the parameters found in Table T1.

This procedure allows us to determine the initial state $\rho(0)$, as well as the Hamiltonian parameters and relaxation/recombination rates. We consider three initial states, all sharing spherical symmetry since the QD is covered by an approximately spherical distribution of chiral bridges:

- a pure singlet state (S), as shown in Figure 4, panel (a);
- a pure triplet state, with the three states equally populated (T), Figure 4, panel (b);
- a mixture of singlet and triplet (ST), with all four levels equally populated, Figure 4, panel (c).

From Figure 4, panels (a,b,c), we see that only the ST case is in qualitative agreement with the measured time dependence, showing a negligible short-time signal that slowly rises until reaching its maximum at $\approx 1.2\ \mu$s. We wish to note that the mixed ST state could be reasonably justified by a small singlet-triplet gap in the precursor state, likely associated with the relatively large size of the QD[59] and a fast inter-multiplet relaxation.[60]

Table T1 reports the Hamiltonian parameters optimized to obtain a better agreement between experiments and theory. For simplicity, we assume an axial dipolar coupling $D = -26$ MHz, which we determined from the separation between different peaks (Figure 5, panels a,e). In turn, by using the point-dipole approximation for calculating $D$ and a distance of 10 Å between QD and $C_{60}$ and neglecting any distribution of conformations for simplicity, we obtain the $g$-value of the QD $g_D \approx 1$. At this distance, one could also expect an isotropic exchange contribution of the same order of magnitude. However, here we prefer to limit the number of parameters and only assume a dipolar contribution. This simplified model leads to $g_A = 2.0037$, higher than literature reports for $C_{60}$ radical anion.[61-62]

As a matter of clarity, it is worth stressing that in Figure 4c, the absence of a signal at short times is due to the equally populated initial CT state. With time, the population of the four levels evolves because of incoherent processes, namely relaxation and recombination. A thorough study of the time dependence of the spectra allows us to fit the relaxation parameters (Table T1). The rate of charge recombination is estimated considering the time at which the maximum signal occurs (Figure 4, panels c,f) and its value of $k_{CR} = 1\ \mu s^{-1}$ is aligned to what we expected for this system.[21] Spin relaxation is described by the two different characteristic times $T_1^D = 200$ ns for the QD and $T_1^A = 10\ \mu$s for $C_{60}$. The former can be estimated by its effect on the relative height of the peaks at times longer than hundreds of nanoseconds, more evident at low temperature (Figure 5, panels a,e), whereas for the latter, we can only infer a lower bound of about $2\ \mu$s, consistently to what expected for the $C_{60}$ anion. An overall dephasing time $T_2 = 33$ ns was obtained by fitting the Lorentzian peak width.

We now consider the possibility of having a CT process that is spin filtered by the CISS effect. To this aim, we simulate the effect of CISS on the trEPR spectra for all three previously considered $\rho(0)$. We model CISS as a "filter" which ideally keeps only the component of the transferred electron spin parallel (or anti-parallel, depending on the enantiomer) to the chiral bridge axis.[18, 63-64] Hence, all the initial states are modified by CISS, but produce different features in the resulting trEPR spectrum on a randomly oriented solution. Notably, we considered only one of the two possible spin orientations because the isotropic QD generates identical EPR spectra for the two enantiomers. As shown in Figure 4d-e, CISS filter strongly affects trEPR spectra for both the singlet and triplet cases, giving rise to opposite *AE* vs. *EA* features at short times. Conversely, CISS does not significantly affect the spectrum with ST initial state (Figure 4e).

Our theoretical analysis suggests that the experimental results are compatible with the presence of CISS in the CT process, although the effect is masked by the ST initial state, which hinders the possibility of distinguishing between a standard CT and a CISS-mediated one. Importantly, our results allow drawing new guidelines for developing model systems for CISS detection. Specifically, our calculations demonstrate that a "pure" precursor (or a precursor with different weights of S and T) would allow unraveling the occurrence of CISS even in a randomly oriented sample. This could occur in systems characterized by a larger singlet-triplet splitting than the CdSe QD employed here. A larger splitting may be achieved, for example, by introducing a shell to increase the electron confinement, as done in literature for similar QD-organic molecule dyads.[21] However, introducing a shell might further reduce the CT efficiency through the long and non-conjugated chiral bridge.

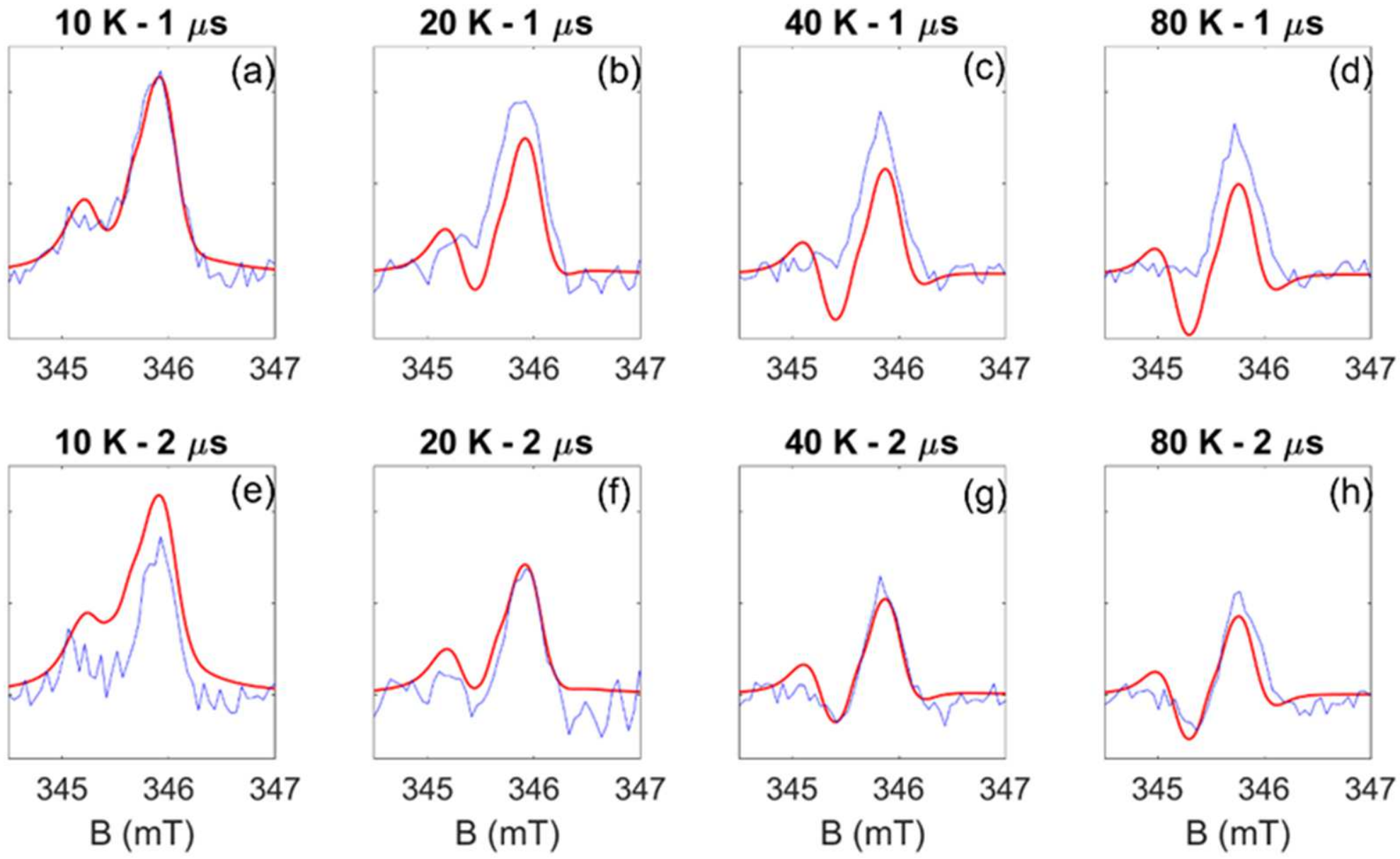

**Figure 5 – Simulated time-resolved EPR spectra as a function of magnetic field, temperature, and time.** Panels (a,b,c,d): spectra calculated at $t$ = 1μs using the four different temperatures $T$=10, 20, 40, 80 K. Panels (e,f,g,h): spectra calculated at $t$ = 2μs at the same temperatures of the top row. All simulations were performed using the parameters in Table T1 and with the ST initial state. The simulations refer to the CISS case. Identical simulations have been obtained for a standard CT without spin filtering due to CISS effect.CT.

**Table T1.** Parameters obtained from the modelling of time-resolved EPR data. D and A indexes refer to the CdSe QD donor and $C_{60}$ acceptor, respectively.

| $g_D$ | $g_A$ | $D$ | $T_1^D$ | $T_1^A$ | $T_2$ | $k_{CR}$ |
|---|---|---|---|---|---|---|
| 1 | 2.0037 | −26 MHz | 200 ns | 10 $\mu$s | 33.4 ns | 1 $\mu s^{-1}$ |

**Conclusions**

In conclusion, we have engineered and developed a model system comprising a CdSe QD as a donor and a $C_{60}$ derivative as an acceptor linked by a rigid and saturated bi-peptidic chiral bridge (χ). This chiral system has shown spin polarization as a result of the photoinduced CT between the CdSe QD and the $C_{60}$ through χ. The CdSe QD-χ-$C_{60}$ system was fabricated through the ligand exchange approach and characterized through the combination of optical spectroscopies and XPS analysis. We then used time-resolved EPR to demonstrate that the photoinduced CT process generates an organic radical localized on the $C_{60}$, which shows a peculiar spin-polarization evolution in the first few μs after the laser pulse. We modeled the trEPR signal of our system in two different cases: (1) a standard spin-polarized photoinduced CT, and (2) a CISS-mediated photoinduced CT. Our calculations demonstrated that the observed EPR features might be compatible with the photoinduced CISS effect.

Though not conclusive yet., our combined experimental and theoretical work represents a first promising attempt toward the direct spectroscopic observation of photodriven CISS effect. Although the search for the perfect model system simultaneously showing good CT efficiencies, efficient spin-filtering, and a well-defined precursor state is still in its infancy, our results suggest that QD-χ-organic molecule dyads are very promising. In addition, our work builds up a sound theoretical framework that will allow a better understanding of spin-polarization arising from photoinduced CT processes in chiral hybrid systems comprising QD and organic molecules.

Ultimately, the possibility of observing a CISS-mediated charge transfer at the molecular level would provide a new tool for molecule-based quantum information processing. Indeed, the QD-χ-$C_{60}$ could form an important building block of a quantum computing architecture, in which polarization resulting from CISS could be harnessed to initialize/readout qubits or implement quantum gates. Thanks to the remarkable efficiency displayed by CISS at high temperatures, this could pave the way toward room-temperature operation of a molecular quantum processor.

## ASSOCIATED CONTENT

### Supporting Information

Synthesis of organic ligands **1**, **2,** and **3**. CdSe Quantum Dots (QDs) fabrication and morphological characterization. Ligand exchange methods. Optical measurements. X-ray photoelectron spectroscopy (XPS) analysis. DFT calculations. Time-resolved Electron Paramagnetic Resonance (trEPR) measurements and best-fit simulations. Theoretical modeling.

## AUTHOR INFORMATION

### Corresponding Authors

*roberta.sessoli@unifi.it

*stefano.carretta@unipr.it

*enrico.salvadori@unito.it

### Notes

The authors declare no competing financial interests.

# ACKNOWLEDGMENT

This work has received funding from the Italian Ministry of Education and Research (MUR) through PRIN Project 2017CR5WCH Q-chiSS ``Quantum detection of chiral-induced spin selectivity at the molecular level" and "Progetto Dipartimenti di Eccellenza 2018-2022 (ref. no. B96C1700020008), and from the European Union's Horizon 2020 research and innovation programme (FET-OPEN project FATMOLS) under grant agreement No 862893. We acknowledge Michael R. Wasielewski and Robert Bittl for their critical manuscript reading.

.

# Supplementary Information for

# Direct Detection of Spin Polarisation in Photoinduced Charge Transfer through a Chiral Bridge

*Alberto Privitera[1,2], Emilio Macaluso[3,4], Alessandro Chiesa[3,4], Alessio Gabbani[5], Davide Faccio[6], Demetra Giuri[6], Matteo Briganti[2], Niccolò Giaconi[2,7], Fabio Santanni[2], Nabila Jarmouni[5], Lorenzo Poggini[8], Matteo Mannini[2], Mario Chiesa[1], Claudia Tomasini[6], Francesco Pineider[5], Enrico Salvadori[1*], Stefano Carretta[3,4*], Roberta Sessoli[2*]*

[1.] *Department of Chemistry, University of Torino, Via Giuria 9, Torino, Italy*

[2.] *Department of Chemistry "U. Schiff" (DICUS), University of Florence, & UdR INSTM Firenze, Via della Lastruccia 3-13, Sesto Fiorentino, Italy*

[3.] *Department of Mathematical, Physical and Information Sciences, University of Parma, I-43124, Parma, Italy & UdR INSTM Parma*

[4.] *INFN–Sezione di Milano-Bicocca, gruppo collegato di Parma, I-43124 Parma, Italy.*

[5.] *Department of Chemistry and Industrial Chemistry, University of Pisa, & UdR INSTM Pisa, Via Moruzzi 13, Pisa, Italy*

[6.] *Department of Chemistry "Giacomo Ciamician", University of Bologna, Via Selmi 2, Bologna, Italy*

[7.] *Department of Industrial Engineering (DIEF), University of Florence, & UdR INSTM Firenze, Via Santa Marta 3, Firenze, Italy*

[8.] *CNR-ICCOM, Via Madonna del Piano 10, Sesto Fiorentino, Italy*

# Table of contents

# DFT calculations

All DFT calculations have been performed with ORCA 4.2.1 quantum chemistry package.[1] For the geometry optimizations, PBE0 functional[2] and D3 empirical dispersion correction[3-4] were used, while def2-TZVP basis set[5] was employed for all the atoms. The thresholds on the maximum force gradient and the energy change were set to $3 \cdot 10^{-4}$ Hartree/Bohr and to $5 \cdot 10^{-6}$ Hartrees, respectively. The root mean square deviation was computed on the heavy atoms C, N, O, and S of the polypeptide chain within the formula $RMSD = \sqrt{\frac{1}{N}\sum_{i=1}^{N} d_i^2}$ where $d_i$ is the distance between a pair of equivalent atoms in the two structures and N is the total number of equivalent atoms. The optimized structure of **1** and the X-ray structure of the S-protected oligopeptide [Tri-S-$(CH_2)_2$-(CO)-(L-Ala-D-Oxd)$_2$-OBn are shown in Figure S13.

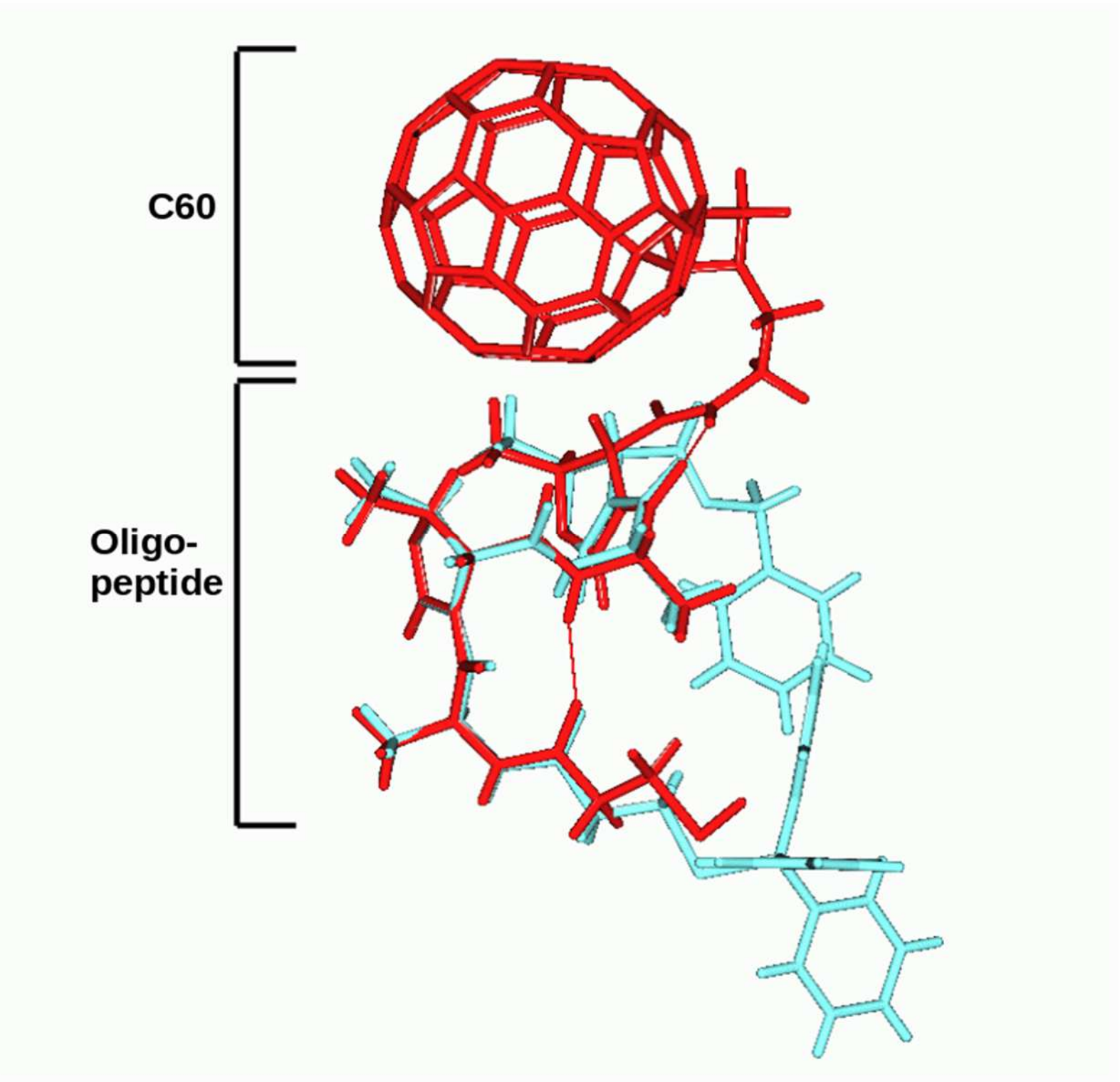

**Figure S1.** Superimposed structures of the optimized geometry of **1** (red) and the x-ray structure of the S-protected oligopeptide lacking fullerene unit (azure blue).

## Synthesis of organic ligands 1, 2 and 3

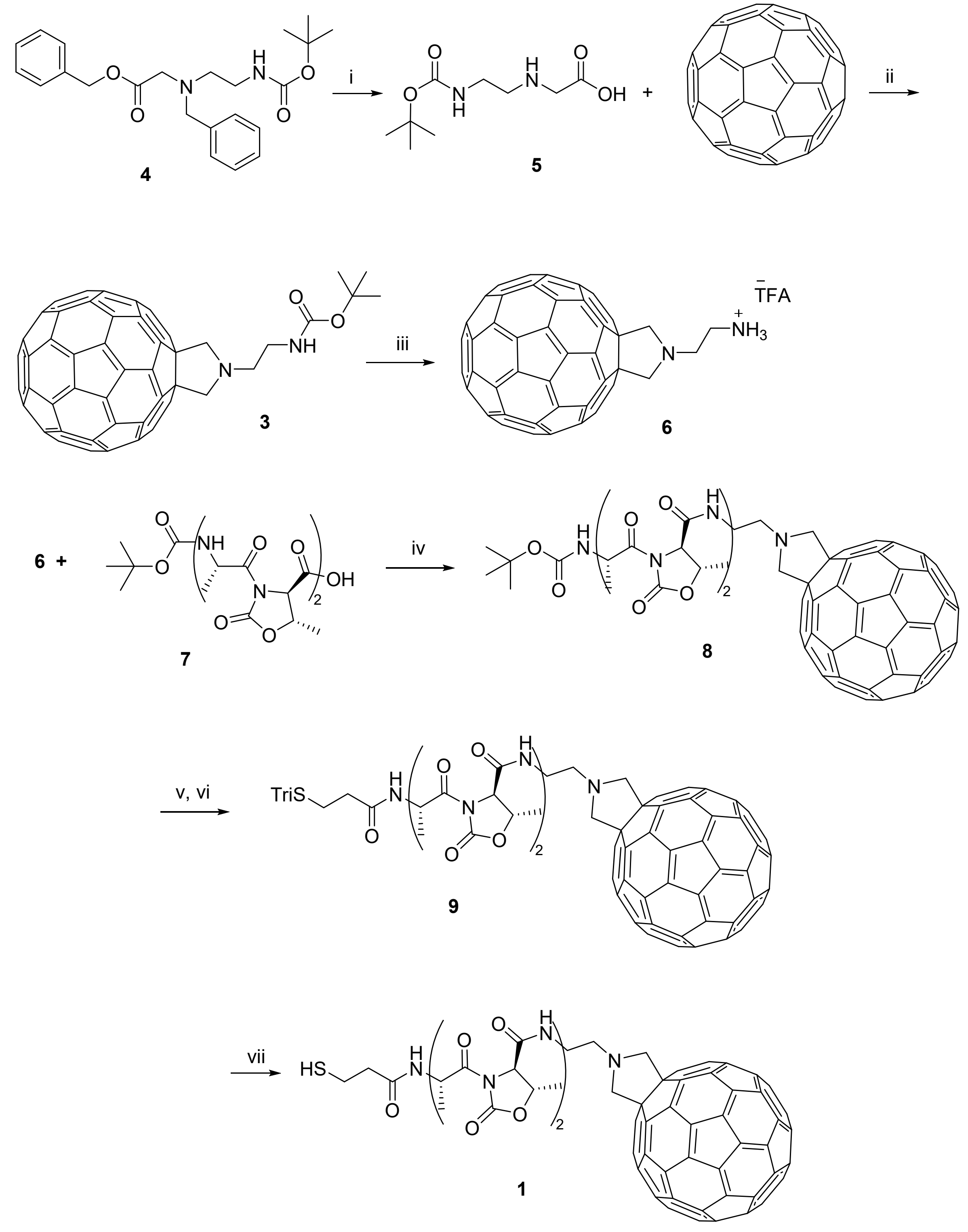

**Scheme S1.** Reagents and Conditions: (i) $H_2$, Pd/C, MeOH, r.t., 4 h, >99% yield; (ii) paraformaldehyde (5 equiv.), toluene, reflux, 1 h, 40% yield; (iii) TFA (200 equiv.), dry $CH_2Cl_2$, r.t., 4 h, >99% yield; (iv) HBTU (2 equiv.), DIEA (2.2 equiv. + some more to remove possible exceeding

TFA from the previous reaction), dry $CH_2Cl_2$, r.t., 24 h, 40% yield; (v) TFA (25 equiv.), $CH_2Cl_2$, r.t., 6 h, 99% yield; (vi) 3-(tritylthio)propanoic acid (1.6 equiv.), HBTU (1.8 equiv.), DIEA (2.2 equiv. + some more to remove possible exceeding TFA from the previous reaction), dry $CH_2Cl_2$, r.t., 24 h, 56% yield; (vi) $Et_3SiH$ (3.8 equiv.) $BF_3.Et_2O$ (0.2 equiv.), HFIP, r.t., 45 min, 60% yield.

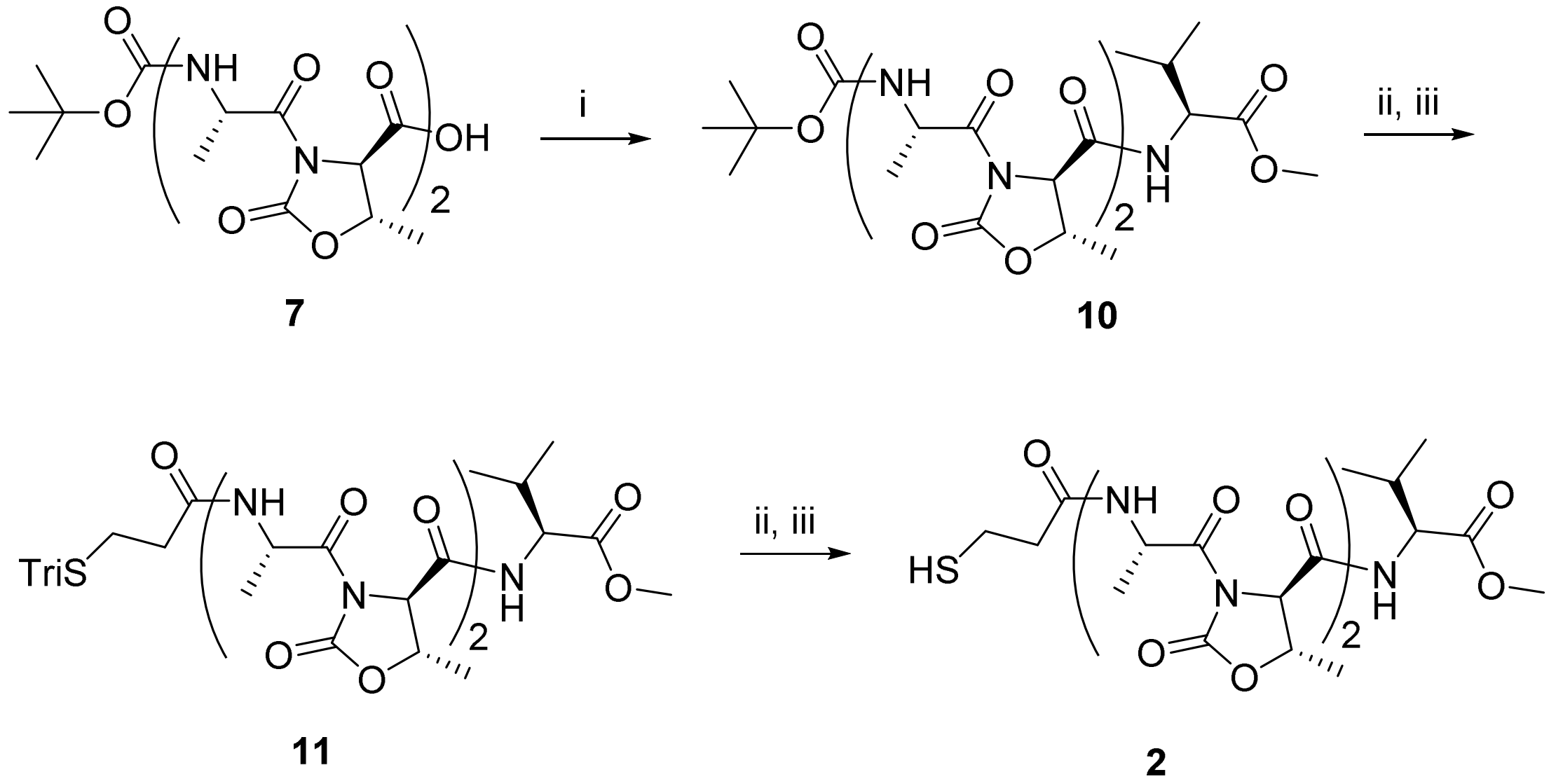

**Scheme S2.** Reagents and Conditions: (i) L-Val-OMe•HCl (1 equiv.), HBTU (1.1 equiv.), DIEA (3.3 equiv.), dry acetonitrile, r.t., 5 h, 90% yield; (ii) TFA (18 equiv.), $CH_2Cl_2$, r.t., 6 h, 99% yield; (iii) 3-(tritylthio)propanoic acid (1 equiv.), HBTU (1.1 equiv.), DIEA (2.2 equiv. + some more to remove possible exceeding TFA from the previous reaction), dry acetonitrile, r.t., 24 h, 92% yield; (iv) $Et_3SiH$ (3.8 equiv.) $BF_3.Et_2O$ (0.2 equiv.), HFIP, r.t., 45 min, 75% yield.

All reactions were carried out in dried glassware and using dry solvents. The melting points of the compounds were determined in open capillaries and are uncorrected. High-quality infrared spectra (64 scans) were obtained at 2 $cm^{-1}$ resolution with an ATR-IR Bruker (Billerica, US, MA) Alpha System spectrometer. All compounds were dried in vacuo, and all the sample preparations were performed in a nitrogen atmosphere. NMR spectra were recorded with a Varian (Palo Alto, US, CA) Inova 400 spectrometer at 400 MHz ($^1$H NMR) and at 100 MHz ($^{13}$C NMR). Chemical shifts are reported in δ values relative to the solvent peak. HPLC-MS analysis was carried out with an Agilent 1260 Infinity II liquid chromatography coupled to an electrospray ionization mass spectrometer (LC-ESI-MS), using a Phenomenex Gemini C18 - 3μ - 110 Å column, $H_2O/CH_3CN$ with 0.2% formic acid as acid solvent at 40 °C (positive ion mode, mz = 50-2000, fragmentor 70 V). To record LC-ESI-MS spectra of the $C_{60}$ derivatives, the compounds were dissolved in 4/1 THF/Methanol.

**Synthesis of 4 and 5.** Benzyl *N*-benzyl-*N*-(2-((*tert*-butoxycarbonyl)amino)ethyl)glycinate **4** was synthesised as reported in literature.[6] The product was then deprotected from the benzyl groups by hydrogenolysis in presence of catalytic Pd/C, and **5** was obtained as a white solid in quantitative yield as reported in literature.[6]

**Synthesis of 3.** The product was obtained following the Prato reaction in 40% yield after silica chromatography purification.[7] The characterization matched the values reported in reference.

**Synthesis of 6.** The Boc group was removed from compound **3** by a 4 h reaction with TFA (trifluoroacetic acid) (200 eq.) in $CH_2Cl_2$, obtaining compound **4** in quantitative yield. The characterization matched the values reported in reference.[7]

**Synthesis of 7.** Boc-(L-Ala-D-Oxd)$_2$-OH (**7**) was synthesized as reported in literature.[8] The characterization matched the values reported in reference.

**Synthesis of 8.** Compound **7** (75 mg, 0.146 mmol) was dissolved in dry $CH_2Cl_2$ (7 mL), then HBTU (2-(1*H*-benzotriazol-1-yl)-1,1,3,3-tetramethyluronium hexafluorophosphate) (63 mg, 0.165 mmol) was added under nitrogen atmosphere and the mixture was stirred at room temperature for 10 minutes. A solution containing **6** (78 mg, 0.097 mmol) and DIEA (98 μL, 0.57 mmol) in dry $CH_2Cl_2$ (3 mL) was added dropwise to the mixture, and the reaction was stirred for 24 h. The mixture was diluted with $H_2O$ and extracted twice with $CH_2Cl_2$. The combined organic layers were dried over $Na_2SO_4$ and the solvent evaporated under vacuum. The crude residue was purified with silica chromatography (99:1 $CH_2Cl_2$:MeOH as solvent). Product **8** was obtained as a brown solid in 40% yield after solvent removal.

IR (ATR-IR): ν 3310, 2921, 2849, 1779, 1686, 1508 $cm^{-1}$. $^1$H-NMR (400MHz, $CDCl_3$): δ 1.40-1.55 (m, 21H, 9H Boc, 2 $CH_3$ Ala, 2 $CH_3$ Oxd), 3.27 (t, 2H, $CH_2$N-pyrrolidine), 3.69-3.80 (m, 2H, C*H*$_2$$CH_2$N-pyrrolidine), 4.45 (m, 4H, $CH_2$ pyrrolidine), 4.47-4.62 (m, 2H, CH α Oxd), 4.70-4.80 (m, 2H, CH α Ala), 5.10 (bs, 1H, NH Boc), 5.17-5.25 (m, 2H, CH β Oxd), 7.19 (1H, NH amide).

**Synthesis of 9.** Product **8** (65.2 mg, 0.05 mmol) was dissolved in dry $CH_2Cl_2$ (2 mL), then TFA (25 equiv., 0.1 mL) was added, and the mixture was stirred for 6 h. When the deprotection was complete, the solvents were removed under reduced pressure and replaced with dry $CH_2Cl_2$ (2 mL). In another flask, 3-(tritylthio)propanoic acid (26 mg, 0.075 mmol) and HBTU (32 mg, 0.085 mmol) were dissolved in dry $CH_2Cl_2$ (6 mL) under nitrogen atmosphere, and the mixture was stirred at room temperature for 10 minutes. The solution containing deprotected **8** and DIEA (*N*,*N*-diisopropylethylamine) (0.106 mL, 0.62 mmol) in dry $CH_2Cl_2$ (2 mL) was then added dropwise. The mixture was stirred for 24 h, then it was diluted with $H_2O$ and extracted three times with $CH_2Cl_2$. The combined organic layers were dried over $Na_2SO_4$, and the solvent evaporated under vacuum. The crude residue was purified with silica chromatography (99:1 $CH_2Cl_2$:MeOH). Product **9** was obtained as a brown solid in 56% yield.

IR (ATR-IR): ν 3205, 2995, 2956, 2916, 2847, 1739, 1697, 1670 $cm^{-1}$. $^1$H-NMR (400MHz, $CDCl_3$): δ 1.27-1.55 (m, 12H, $CH_3$-Ala, $CH_3$-Oxd), 2.44-2.56 (m, 2H, *CH$_2$*$CH_2$STrt), 2.77-3.05 (m, 2H, $CH_2$*CH$_2$*STrt), 3.20 (m, 2H, NH $CH_2$*CH$_2$*N Fulleropyrrolidine), 3.68-3.77 (m, 4H, 2 CH-α-Oxd, NH*CH$_2$*$CH_2$N Fulleropyrrolidine), 4.25 (q, 1H, CHα-Ala) 4.34-4.50 (m, 5H, 2 $CH_2$ Fulleropyrrolidine, 1 CHα-Ala), 4.54-4.75 (m, 2H, CHβ-Oxd), 7.13-7.43 (m, 15 H, CH Ph).

**Synthesis of 1.** Compound **8** (40 mg, 0.028 mmol) was added to a solution of $BF_3$•$Et_2O$ (1 μL, 8.4 μmol), $Et_3SiH$ (0.017 mL, 0.107 mmol) in hexafluoroisopropanol (HFIP 0.15 M, 0.112 mL), following a procedure reported in the literature.[9] After 45 min under stirring, the reaction was quenched with $NaHCO_3$ and the residue was extracted three times with $CH_2Cl_2$. The combined organic layers were dried over $Na_2SO_4$ and the solvent was evaporated under vacuum. The crude residue was purified with silica chromatography (98:2 $CH_2Cl_2$:MeOH). Product **1** was obtained as a brown solid (yield 60%). IR (ATR-IR): ν 3060, 3022, 2922, 2855, 1782, 1732, 1652, 1595, 1494 $cm^{-1}$. $^1$H-NMR (400MHz, $CDCl_3$): δ 1.21-1.7 (m, 13H, $CH_3$-Ala, $CH_3$-Oxd, SH), 2.51 (t, 2H, $CH_2$*CH$_2$*SH), 2.63 (t, 2H, *CH$_2$*$CH_2$SH), 3.25 (m, 2H, NH $CH_2$*CH$_2$*N Fulleropyrrolidine), 3.70-3.74 (2H, NH*CH$_2$*$CH_2$N Fulleropyrrolidine), 3.76-3.83 (m, 2H, CH-α-Oxd), 4.24 (q, 1H, CH-α-Ala),

4.39-4.52 (m, 5H, 2 $CH_2$ Fulleropyrrolidine, 1 CH-α-Ala), 4.56-4.76 (m, 2H, CH-β-Oxd). $^{13}C$-NMR (400 MHz, DMSO, *d₆*): δ 170.6, 158.6, 155.9, 147.2, 146.6, 146.2, 145.7, 145.2, 145.2, 144.8, 144.5, 143.0, 142.5, 142.3, 142.0, 141.8, 139.9, 136.4, 136.2, 129.5, 128.5, 127.6, 127.16, 76.4, 71.2, 67.38, 62.44, 61.4, 60.2, 55.3, 53.3, 38.6, 26.0, 21.4, 21.2, 18.4, 17.1, 14.5. HPLC-MS (ESI): 12.44 min, [M+H$^+$] =1292.

**Synthesis of 10.** Compound **7** (170 mg, 0.33 mmol) was dissolved in dry $CH_3CN$ (7 mL). Then HBTU (138 mg, 0.364 mmol) was added under nitrogen atmosphere. The mixture was stirred at room temperature for 10 minutes. A solution containing L-Val-OMe•HCl (55 mg, 0.33 mmol) and DIEA (0.185 mL, 1.09 mmol) in dry $CH_3CN$ (3 mL) was then added dropwise, and the mixture was stirred for 5 h. The solvent was evaporated under vacuum. The residue was dissolved in $CH_2Cl_2$ and washed with distilled $H_2O$, 1M HCl, aqueous $NaHCO_3$ and $H_2O$. The organic layers were dried over $Na_2SO_4$, and the solvent was evaporated under vacuum. The crude residue was washed with $Et_2O$ and *n*-hexane and filtered. Product **10** was obtained as a white solid with a yield of 90%.

M.p. 106-108 °C; IR (ATR-IR): ν 2979, 1784, 1747, 1733, 1717, 1701, 1681, 1672, 1653, 1557 $cm^{-1}$.$^1$H NMR ($CDCl_3$, 400 MHz): δ 0.92 (d, 6H, 2 $CH_3$-Val), 1.39 (d, 3H, $CH_3$-Oxd), 1.41 (9H, Boc), 1.49 (d, 3H, $CH_3$-Oxd), 1.51 (d, 3H, $CH_3$-Ala), 1.53 (d, 3H, $CH_3$-Ala) 2.15 (m, 1H, CHβ-Val), 3.70 (s, 3H, Val-OMe), 4.42 (d, 1H, CHα-Val), 4.46 (q, 2H, CHα-Ala), 4.69-4.81 (m, 2H, CHα-Oxd), 5.05 (d,1H, NH-Boc), 5.19-5.38 (m, 2H, CHβ-Oxd), 7.13 (d, 1H, NH amide), 7.34 (bs, 1H, NH amide). HPLC-MS (ESI): 8.08 min, [M+Na$^+$] = 650, [(M-Boc)+H$^+$] = 528.

**Synthesis of 11.** Product **10** (65.2 mg, 0.05 mmol) was dissolved in dry $CH_2Cl_2$ (2 mL), then TFA (18 equiv., 70 μL) was added, and the mixture was stirred for 4 h. When the deprotection was complete, the solvents were removed under reduced pressure and replaced with dry $CH_2Cl_2$ (2 mL). In another flask, 3-(tritylthio)propanoic acid (115 mg, 0.33 mmol) and HBTU (138 mg, 0.364 mmol) were dissolved in dry $CH_2Cl_2$ (6 mL) under nitrogen atmosphere. The mixture was stirred at room temperature for 10 minutes. The solution containing deprotected **8** and DIEA (*N,N*-diisopropylethylamine) (0.125 mL, 0.726 mmol) in dry $CH_2Cl_2$ (2 mL) was then added dropwise. The mixture was stirred for 5 h, then it was diluted with $H_2O$ and extracted three times with $CH_2Cl_2$. The combined organic layers were dried over $Na_2SO_4$, and the solvent evaporated under vacuum. Product **12** was obtained as a yellowish solid after washing with $Et_2O$ and *n*-hexane with a yield of 92%.

M.p. 158-160 °C. IR (ATR-IR): ν 3280, 3056, 2963, 2933, 1777, 1740, 1707, 1654, 1543 $cm^{-1}$. $^{1}H$ NMR ($CDCl_3$, 400 MHz): δ 0.91 (m, 6H, $CH_3$-Val), 1.31-1.40 (m, 6H, $CH_3$-Oxd), 1.46 (d, 3H, $CH_3$-Ala), 1.52 (d, 3H, $CH_3$-Ala), 1.93-2.24 (m, 3H, $CHCH(CH_3)_2$-Val, $TrtSCH_2CH_2CO$), 2.32-2.53 (m, 2H, $SCH_2CH_2CO$), 3.70 (s, 3H, $OCH_3$), 4.37-4.48 (m, 3H, CHα-Val, 2 CHα-Oxd), 4.59 (m, 1H, CHβ-Oxd), 4.76 (m, 1H, CHβ-Oxd), 5.24 (m, 2H, CHα-Ala), 6.26 (bs, 1H, NH), 7.15-7.43 (m, 15H, CH-Trt), 7.72 (bs, 1H, NH). HPLC-MS (ESI): 11.21 min, $[M+Na^+]$ = 880.

**Synthesis of 2.** *S*-detritylation of **12** was obtained as reported in literature and previously described.[9] The residue was washed with *n*-hexane, and product **2** was obtained as a white solid in yield 75%.

M.p. 175-178 °C. IR (ATR-IR): ν 3273, 3065, 2963, 1779, 1739, 1706, 1652, 1537, 1447 $cm^{-1}$. $^{1}H$ NMR ($CDCl_3$, 400 MHz): δ 0.92 (m, 6H, $CH_3$-Val), 1.45-1.55 (m, 13H, $CH_3$-Ala, $CH_3$-Oxd, SH), 2.15 (m, 1H, $CHCH(CH_3)_2$-Val), 2.55 (m, 2H, $HSCH_2CH_2CO$), 2.75 (m, 2H, $SCH_2CH_2CO$), 3.71 (s, 3H, $OCH_3$), 4.42 (m, 1H, CHα-Val), 4.47 (m, 2H, CHα-Oxd), 4.68 (m, CHβ-Oxd), 4.79 (m, CHβ-Oxd), 5.36 (m, 2H, CHα-Ala), 7.12 (1H, NH), 7.39 (1H, NH), 7.63 (1H, NH). $^{13}C$ NMR ($CDCl_3$, 400 MHz): δ 173.6, 172.94, 171.9, 171.5, 167.5, 167.3, 151.8, 151.7, 75.2, 74.9, 62.7, 62.3, 58.1, 52.1, 49.1, 48.8, 39.5, 30.6, 21.4, 21.2, 20.1, 18.9, 18.6, 15.7. HPLC-MS (ESI): 6.77 min, $[M+H^+]$ = 616, $[M+Na^+]$ = 638.

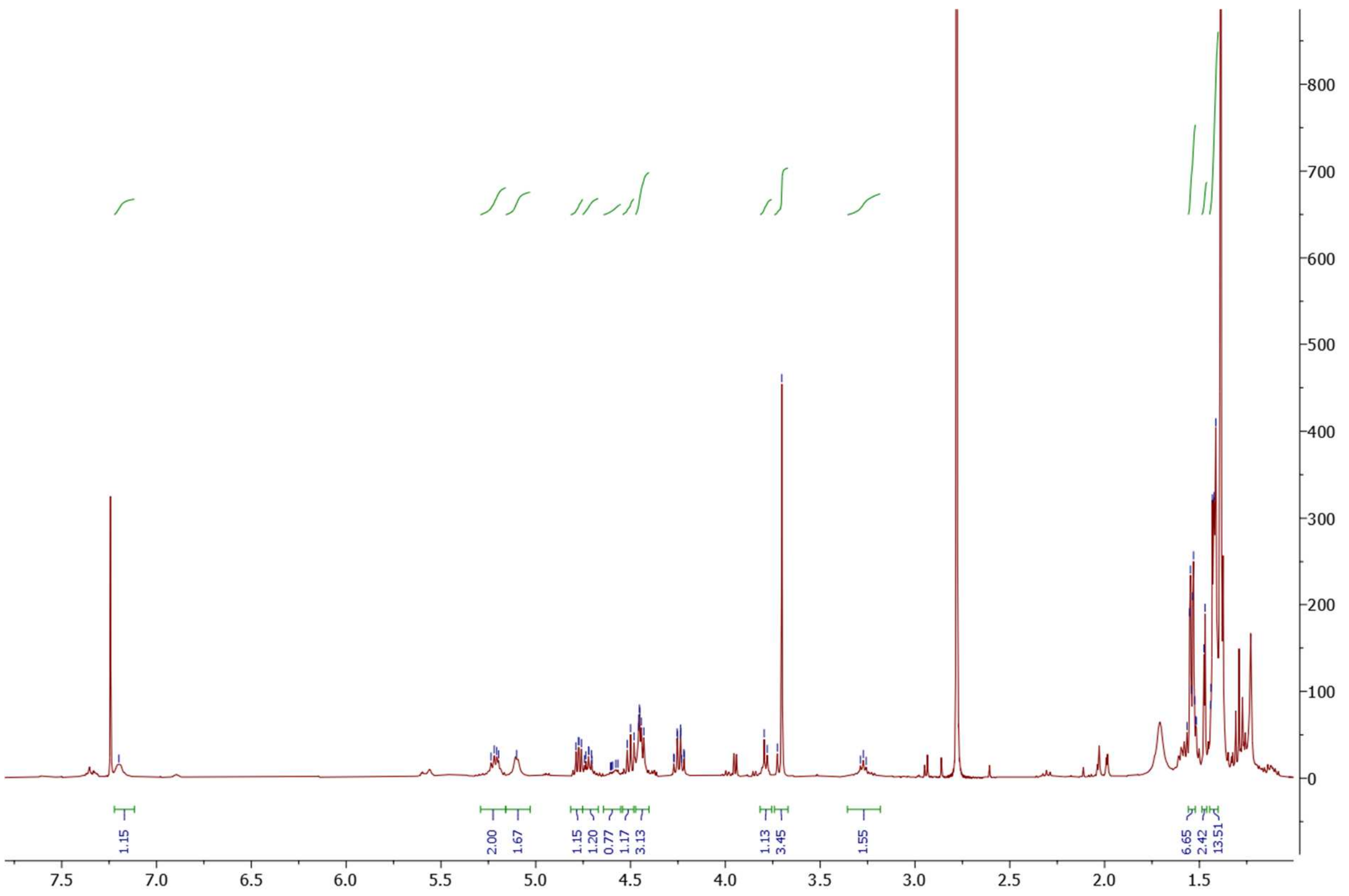

**Figure S2.** $^{1}$H NMR spectrum of Product **8.**

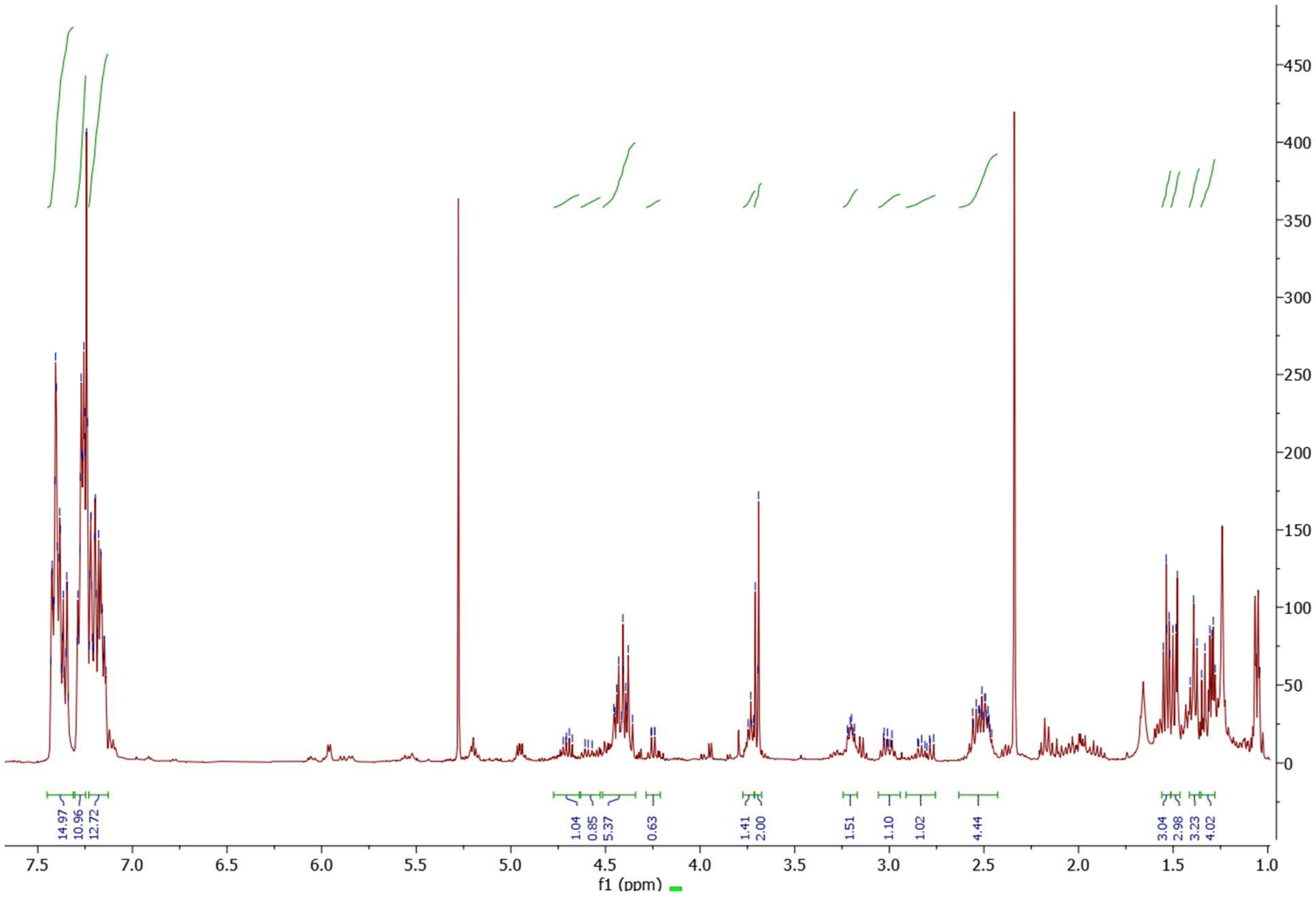

**Figure S3.** $^{1}$H NMR spectrum of Product **9.**

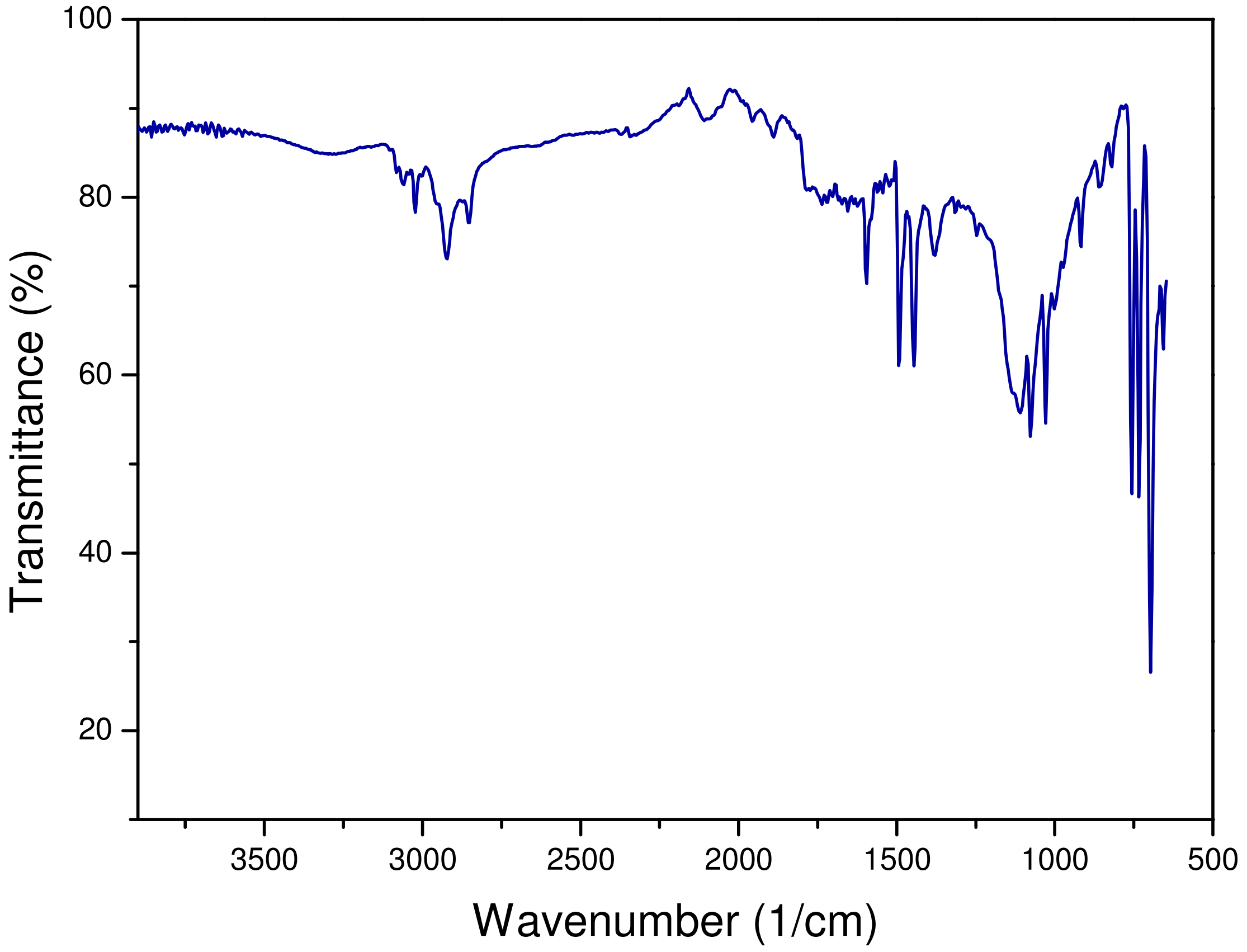


**Figure S4.** ATR-IR spectrum of product **1.**

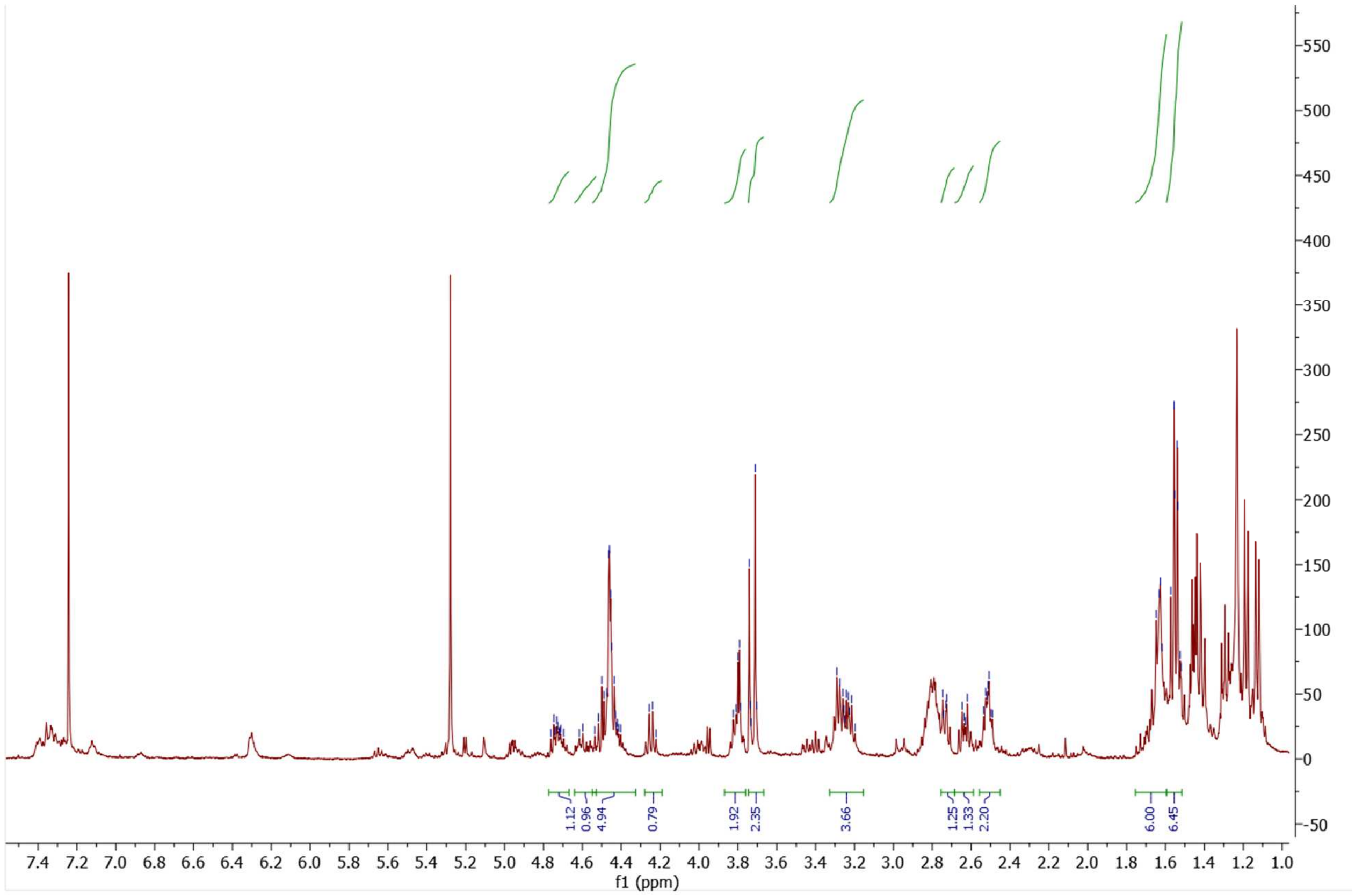

**Figure S5.** $^{1}$H NMR spectrum of Product **1**

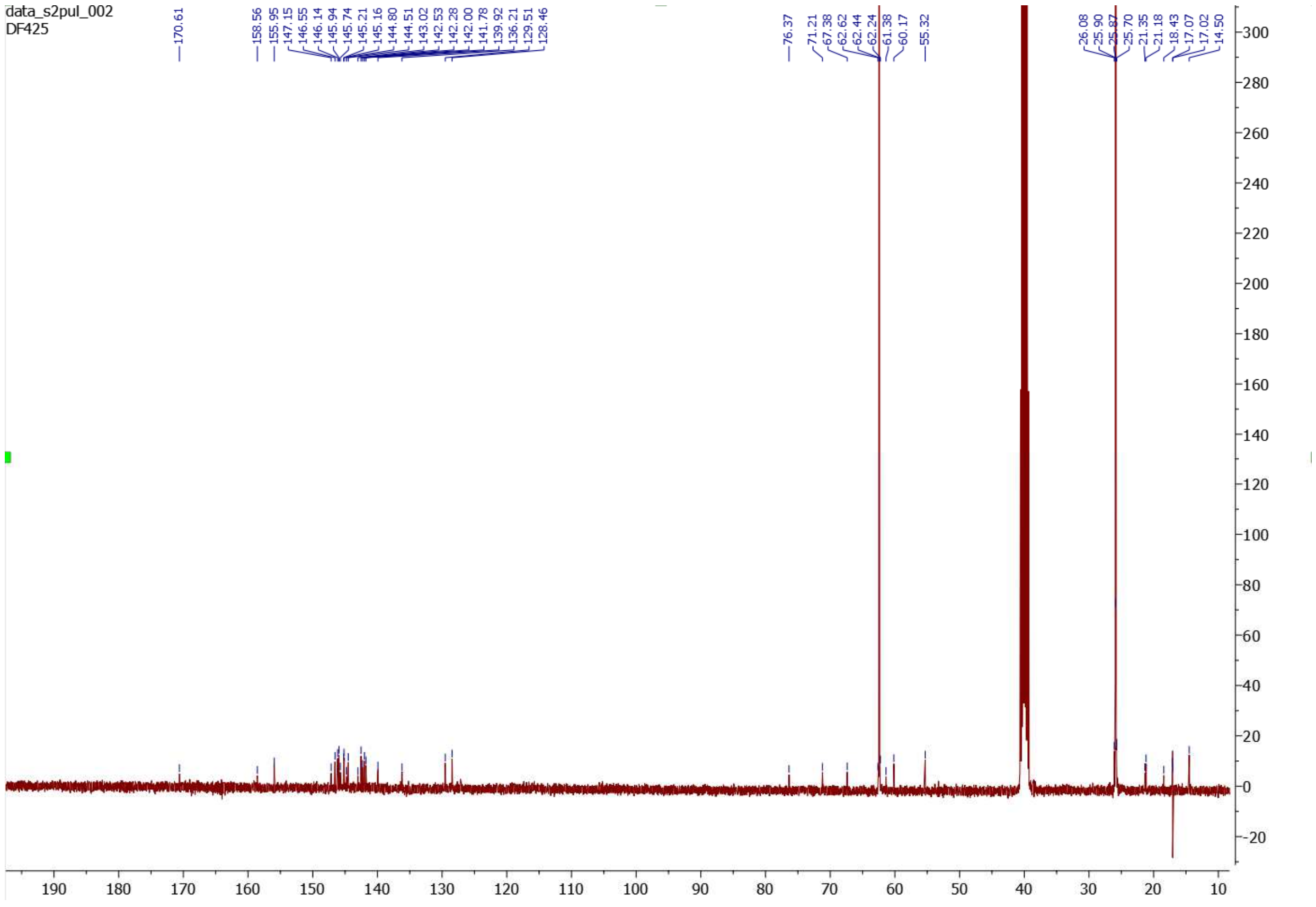

**Figure S6.** $^{13}$C NMR spectrum of Product **1.**

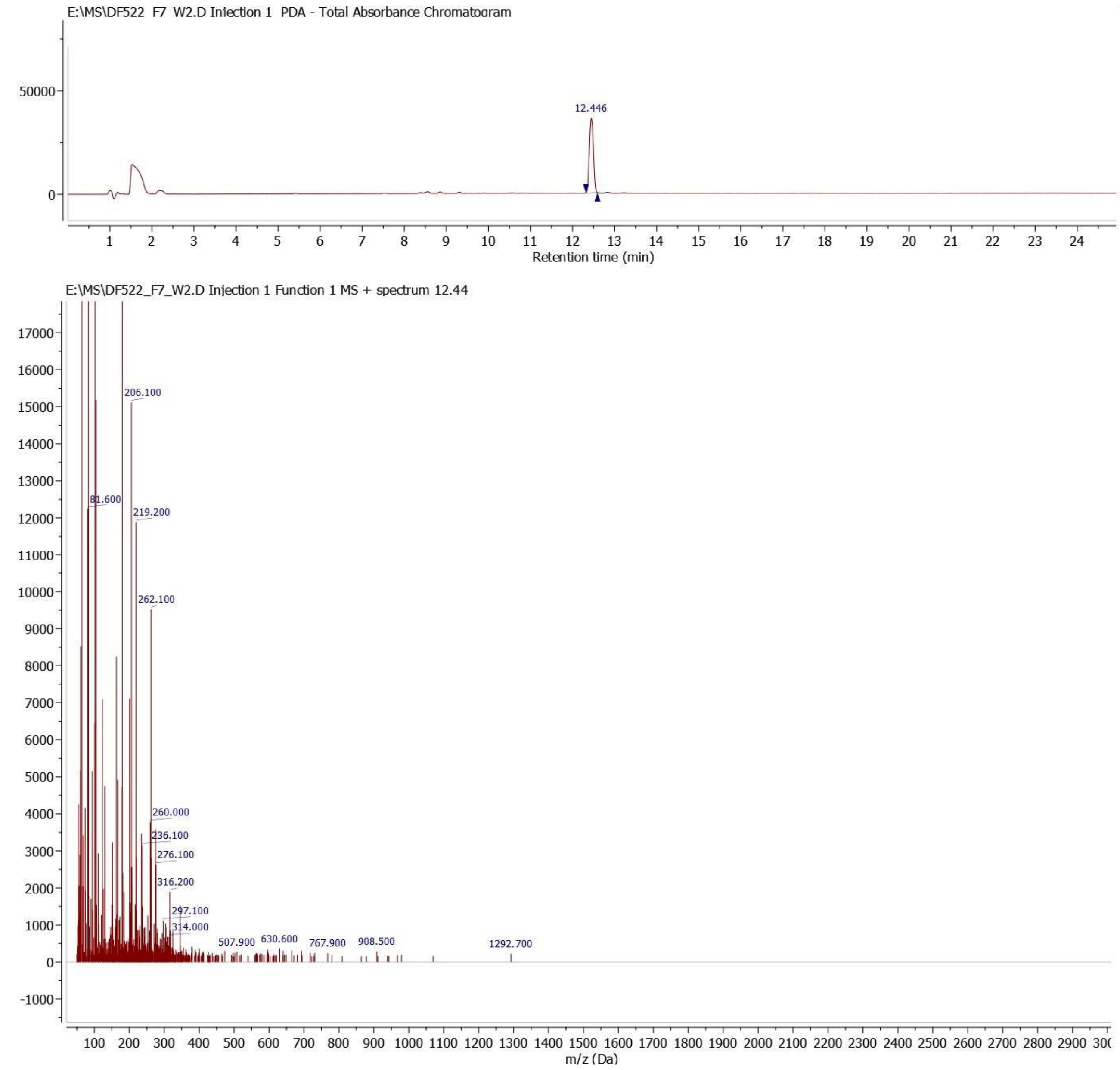


**Figure S7.** HPLC-MS analysis of product **1.**

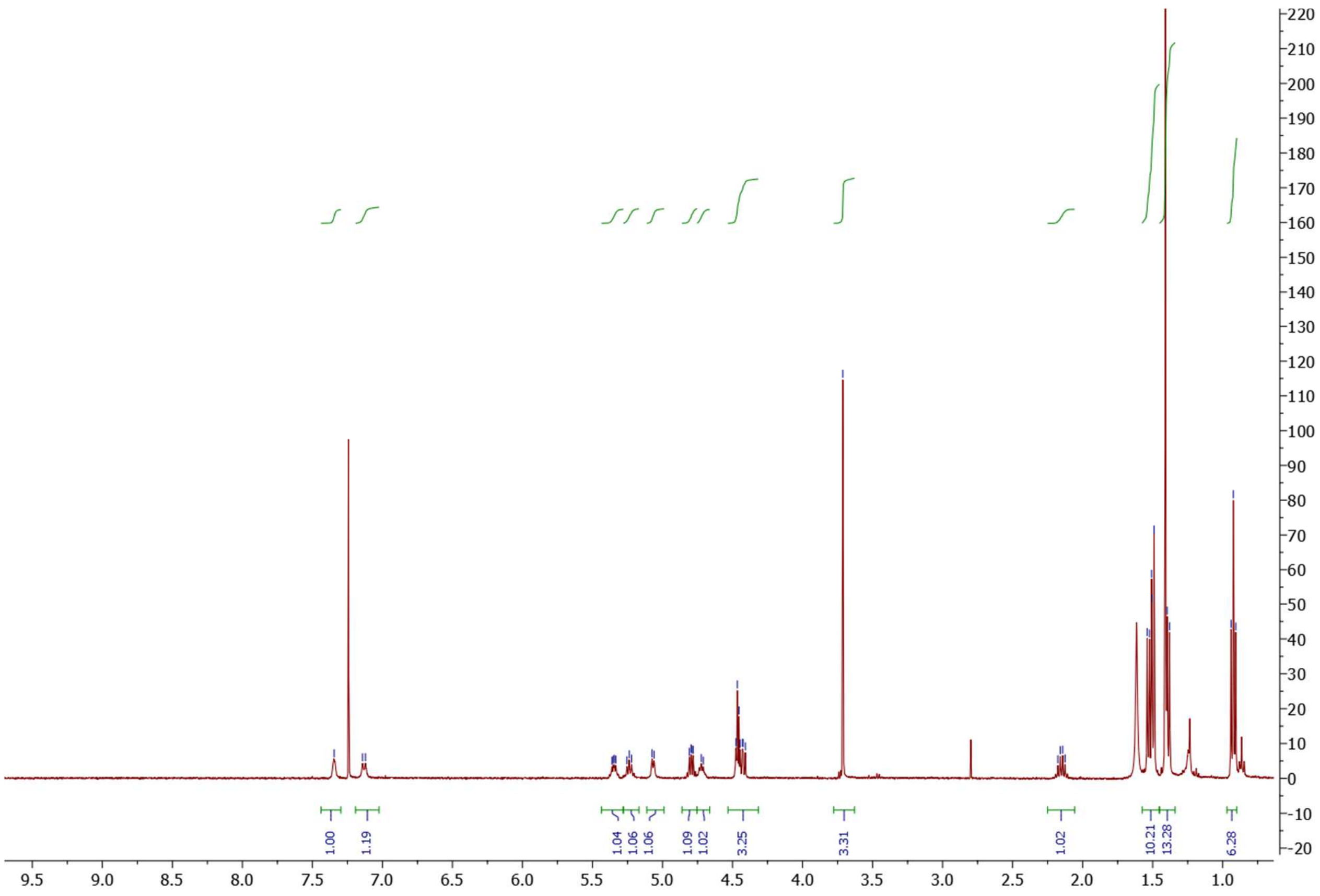


**Figure S8.** $^{1}$H NMR spectrum of Product **10.**

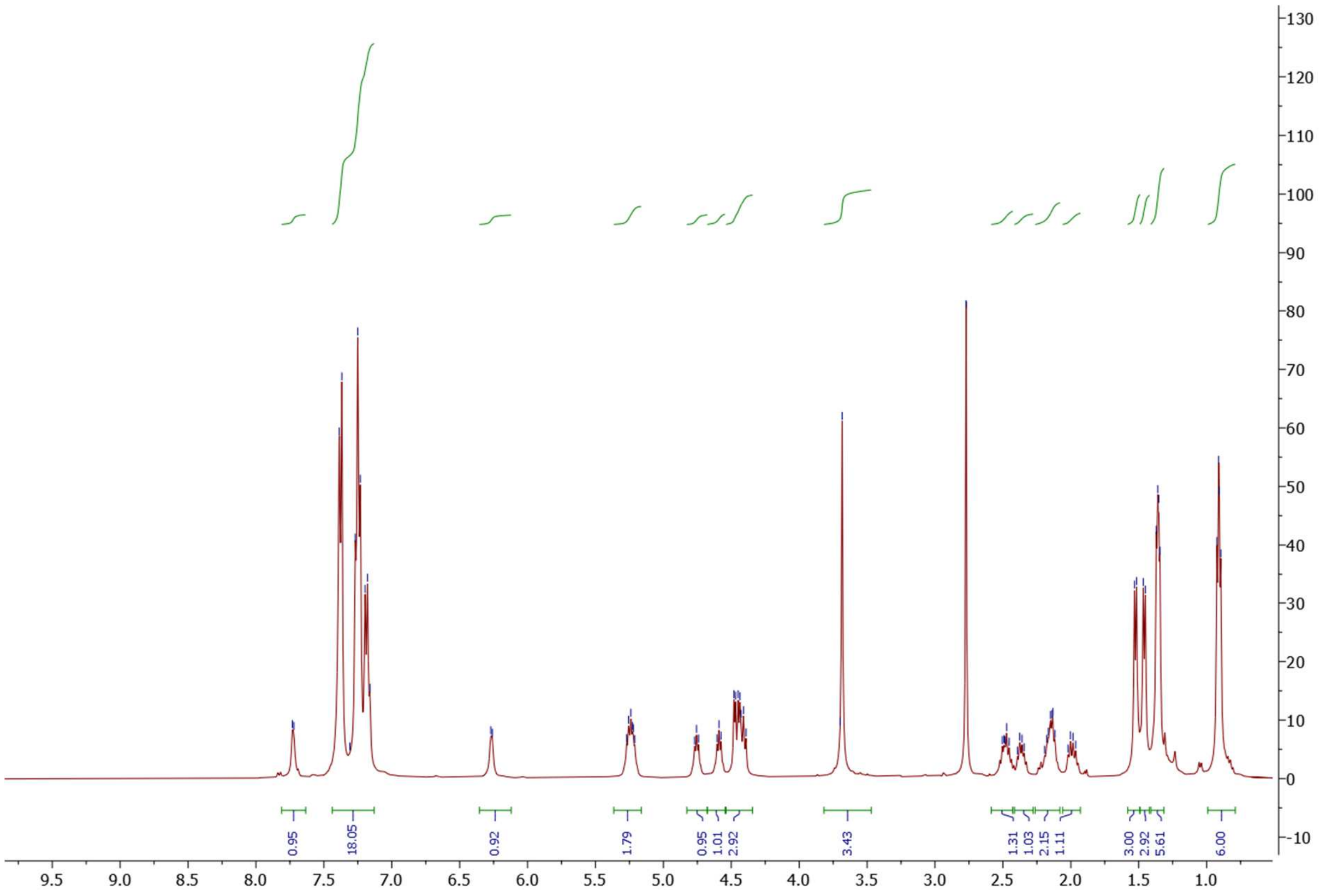


**Figure S9.** $^{1}$H NMR spectrum of Product **11.**

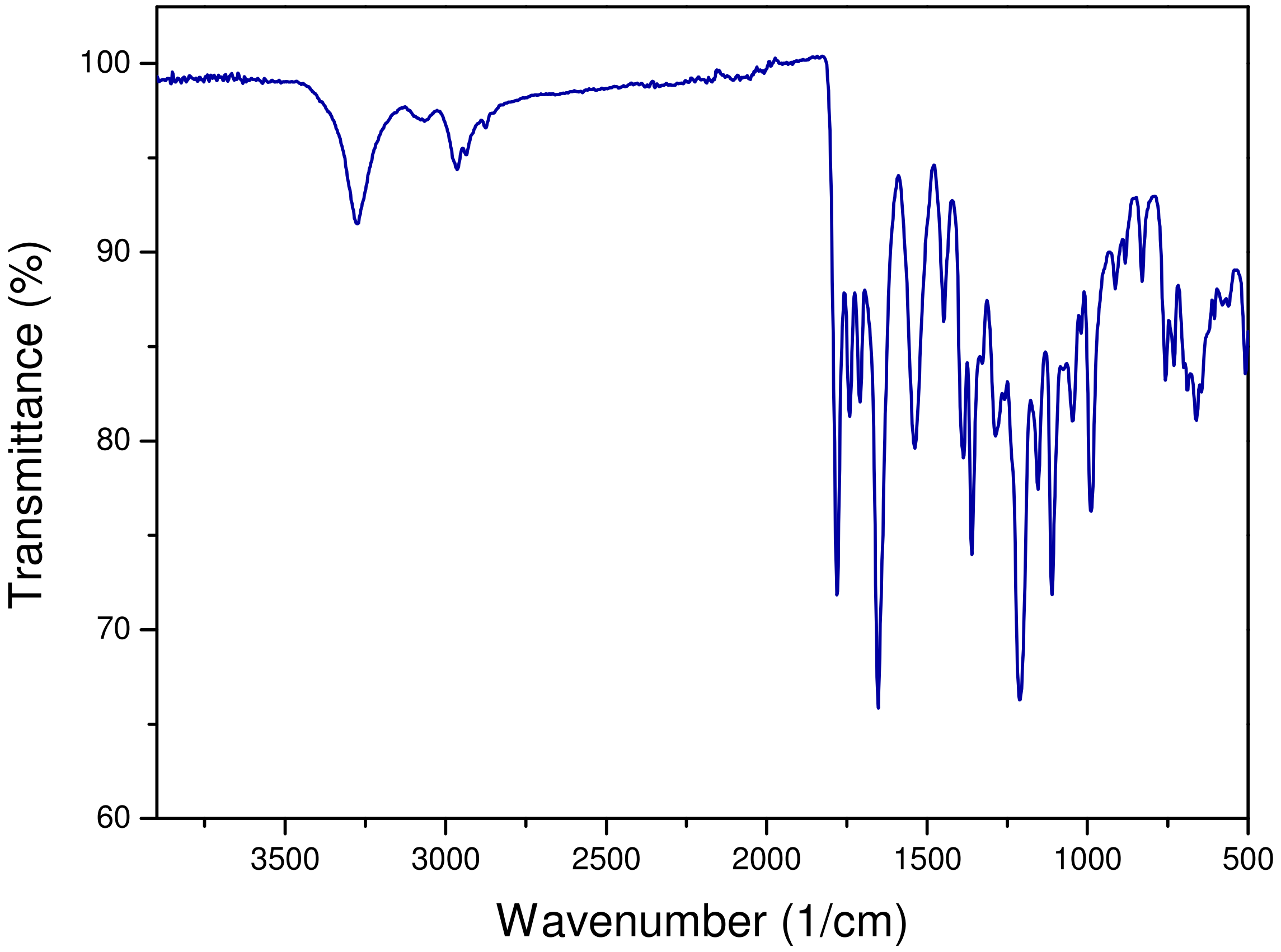


**Figure S10.** ATR-IR spectrum of product **2.**

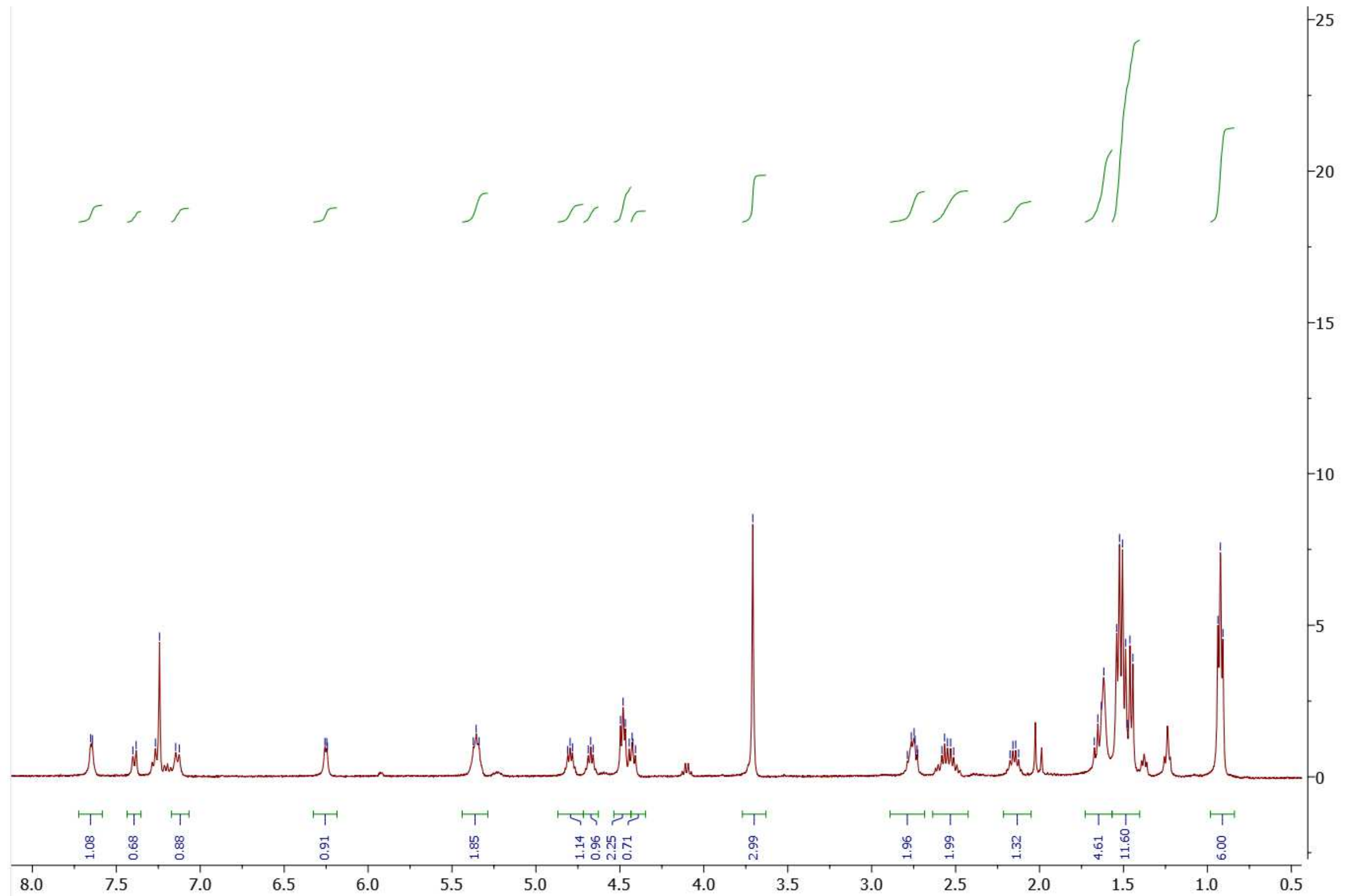


**Figure S11.** $^{1}$H NMR spectrum of Product **2.**

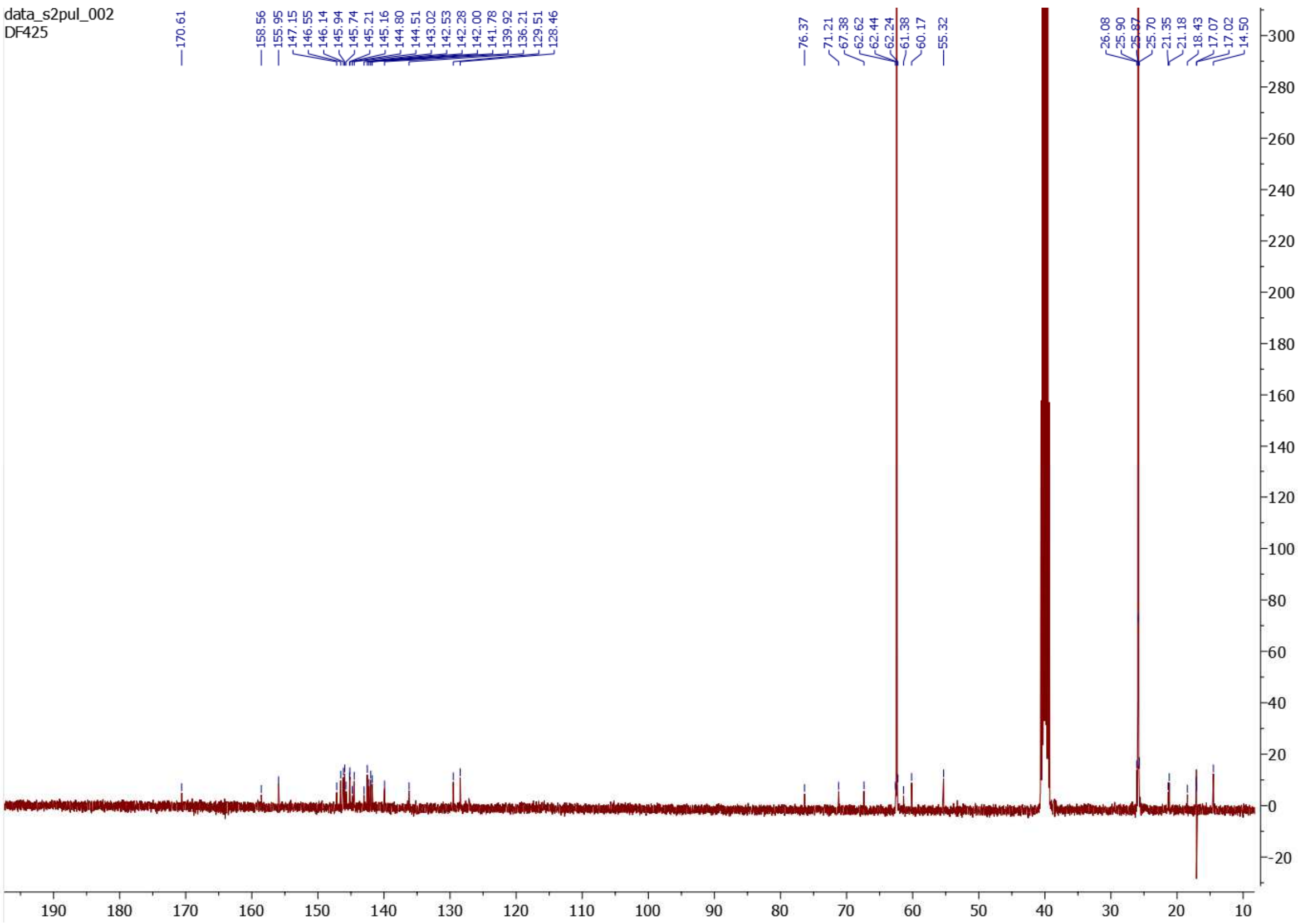


**Figure S12.** $^{13}$C NMR spectrum of Product **2.**

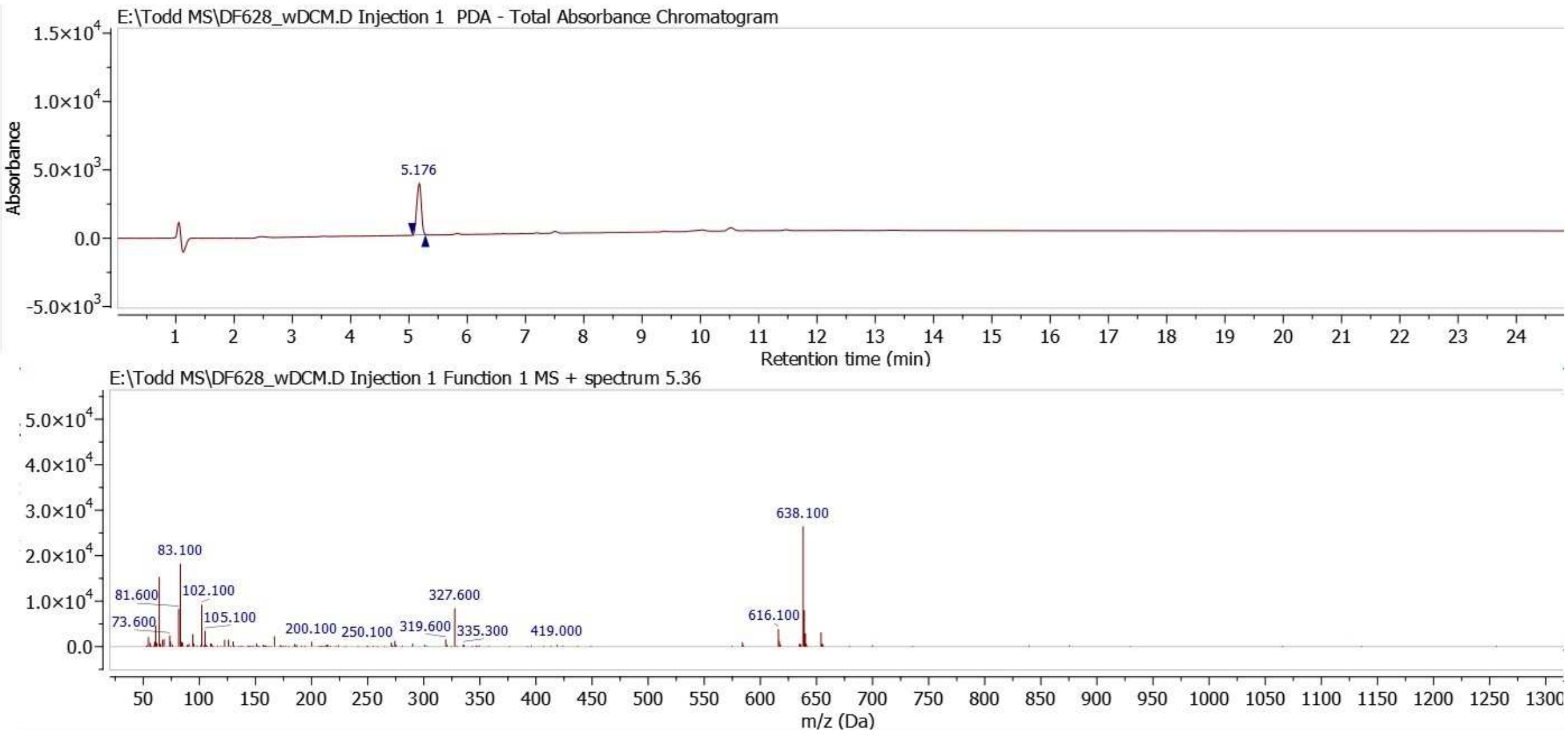


**Figure S13.** HPLC-MS analysis of product **2**

# CdSe Quantum Dots (QDs) fabrication

### Synthesis of pristine CdSe QDs

The CdSe QDs were prepared according to the hot injection method reported by Quanqin Dai *et al* with minor modifications.[10] Specifically, 0.473 mmol of CdO, 4.8 mmol of oleic acid, 14 mL of 1-octadecene, and 9.4 mmol of trioctyl phosphine oxide (TOPO) were loaded into a three-neck flask, and heated to 300°C under $N_2$ atmosphere. At this temperature, a TOP-Se solution (prepared by adding 1.7 mL of TOP to 0.93 mmol of Se powder in 4.1 mL of ODE) was swiftly injected. After the injection, the temperature was decreased to 270°C, and the CdSe QDs were let grow for 50 min. The reaction was quenched by removing the heating source and adding toluene. Finally, the CdSe QDs were purified by four washing steps centrifugating in ethanol at 4500 *g*, where *g* is the gravitational acceleration, and redispersing the precipitate in toluene, and finally dispersed in chloroform.

### Morphological characterization

TOPO-capped QDs were characterized using Transmission Electron Microscopy (TEM) and Powder X-ray Diffraction (PXRD). TEM analysis was performed using a JEOL 100 SX, operating at 100 kV. Samples were prepared by drop casting a dilute suspension of CdSe QDs in hexane onto 200 mesh carbon-coated copper grids. The mean size and size distribution were obtained from statistical analysis over at least 300 CdSe QDs using the image J software.[11] Powder X-ray Diffraction (PXRD) measurements were carried out using a Bruker D8 Advance diffractometer equipped with a Cu Kα radiation and operating in Theta-Theta Bragg Brentano geometry at 40 kV and 40 mA.

TEM analysis reveals a size distribution of 5 nm with a standard deviation of 1 nm (Figure S1). PXRD measurements reveal broad diffraction peaks typical of QDs with similar size (Figure S2). The diffraction pattern is consistent with CdSe phases, while the reduced size of the crystallites makes difficult to univocally distinguish from cubic zinc blend or wurtzite crystal phases. This is typically reported for chalcogenide CdX (with X=S, Se, Te) QDs with size below 10 nm.[12]

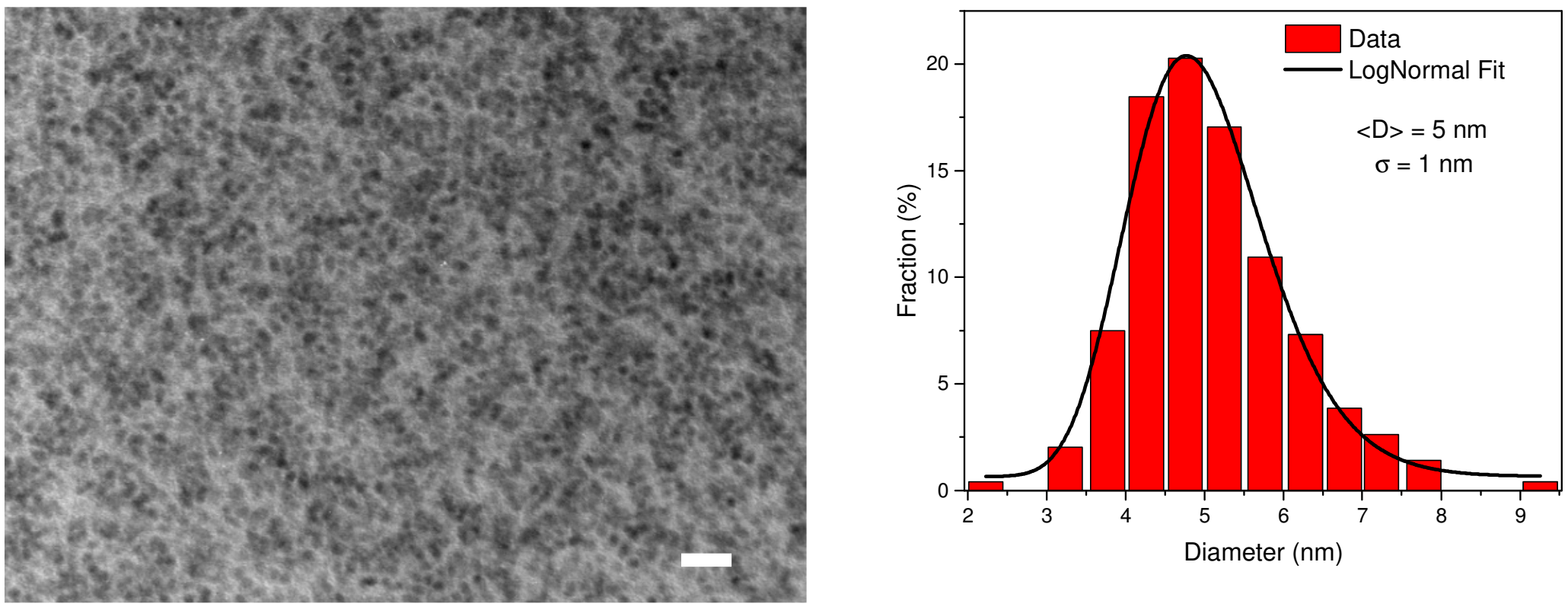


**Figure S14: a)** Representative TEM image of as-synthesized CdSe QDs (scale bar is 20 nm); b) size distribution obtained from statistical analysis of TEM images. The black line represents the fitting with a LogNormal distribution function. The average size <D> and standard deviation σ are displayed in the graph.

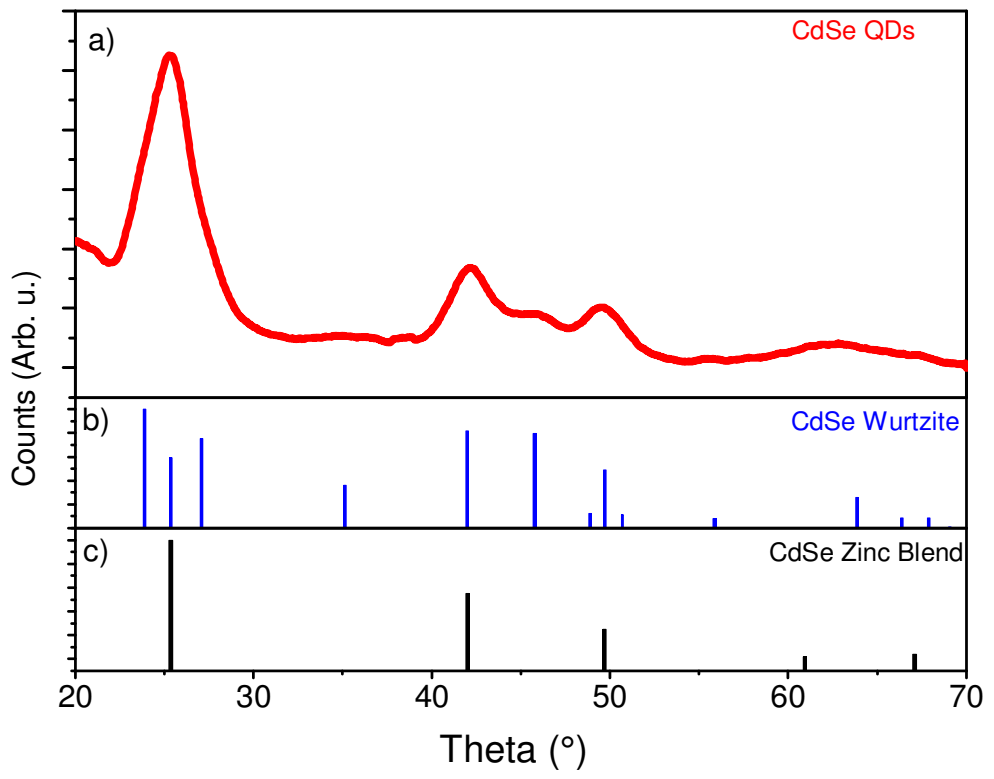


**Figure S15:** a) Diffraction pattern of as-synthesized CdSe QDs; b-c) reference pattern of the wurtzite (Powder Diffraction File: PDF 65-3415) and zinc blend (Powder Diffraction File: PDF 19-0191) CdSe crystal phases.

**Ligand exchange**

8.5 mg of **1** were added to a solution containing 1.6 mg of TOPO-capped CdSe QDs in chloroform. The **1**-to-CdSe ratio was chosen considering a 10x molar excess with respect to the maximum number of molecules that can be adsorbed on the QDs surface considering a maximum packing density on the basis of the structure of the molecule. The mixture was kept under constant stirring overnight at room temperature. At the end of the ligand exchange process, the formed precipitate was observed and washed three times by centrifugation in ethanol. Absorption and luminescence spectra confirm the presence of the nanoparticles in the precipitate. Following a solubility test in different solvents (toluene, chloroform, dichloromethane, 1,2,4-trichlorobenzene), the **1**-capped QDs were finally dispersed in 1,2,4-trichlorobenzene.

A similar procedure was carried out to synthesise the control sample consisting of CdSe QDs ligand-exchanged with **2**.

# Optical spectroscopy

## Methods

UV-Vis spectra was recorded at room temperature by using a Cary 5000 spectrophotometer (Agilent) in the range 400-800. The PL spectra were measured at room temperature using an excitation wavelength ($\lambda_{ex}$) of 400 nm on a FluoroMax P (Horiba). The measurements were performed using quartz cuvettes with a path length of 1 cm. PL spectra were normalized for the optical density of the same solution at the maximum of the excitonic peak.

Time-resolved photoluminescence (TRPL) measurements were carried out using the time-correlated single-photon counting (TCSPC) technique. The experimental apparatus is based on the FluoroMax P spectrofluorometer detection unit (grating monochromator and photomultiplier tube), powered by the FluoroHub Single Photon Counting unit. The excitation source was a blue pulsed Horiba NanoLED, generating picosecond pulses in the UV (375 nm). The instrument response function (IRF) for the whole apparatus was determined by means of scattered light detection, by using a reference sample of LUDOX® colloidal silica. LED radiation was focused by means of a spherical lens on a sample holder and sample emission was collected with a 90° geometry to minimize scattering interferences. The measurements were carried out on dilute solutions of TOPO-capped and **1**-capped QDs solutions in 1,2,4 trichlorobenzene. Solutions were prepared keeping optical absorbance at 400 nm below 0.2 in 1 cm path quartz cuvettes.

## Optical measurements

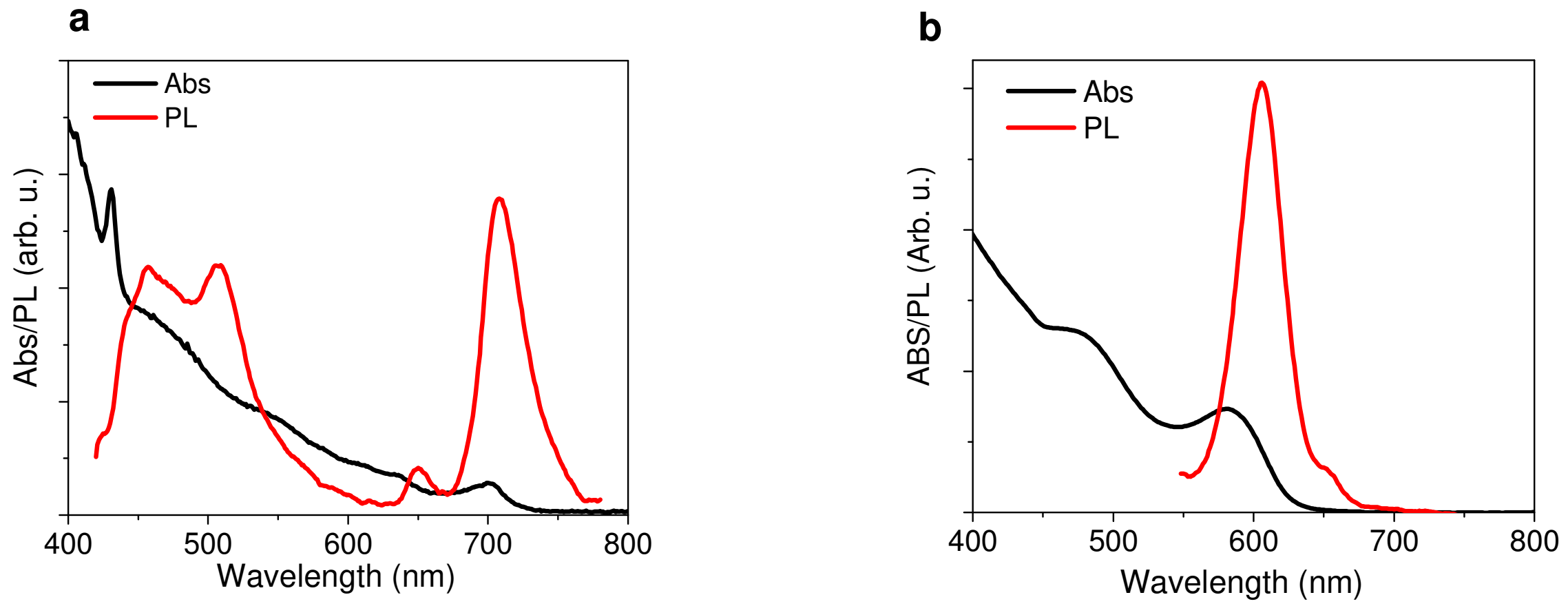


**Figure S16:** Absorption and photoluminescence spectra of (a) **1** and (b) the CdSe QDs functionalized with **2** (QDs-χ).

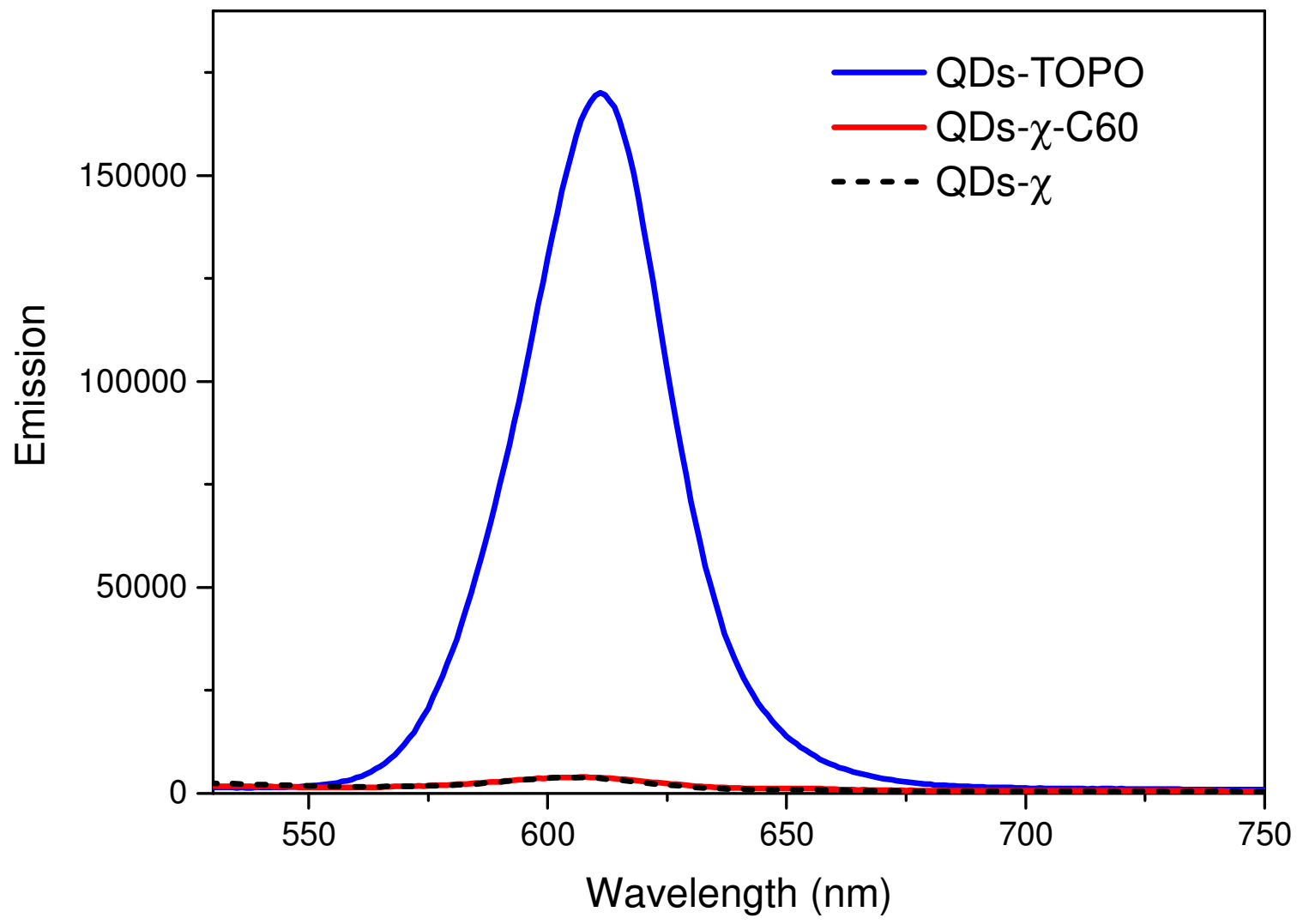


**Figure S17**: Comparison of the emission intensity of the CdSe QDs functionalized with TOPO (QDs-TOPO), **1** (QDs-χ–C60) and **2** (QDs-χ) dispersed in 1,2,4-trichlorobenzene at the same concentration.

| | **QD-χ-C60** | **CdSe QDs** |
|---|---|---|
| **$\tau_1$ ($a_1$)** | 0.07 (0.99) | 3.4 (0.65) |
| **$\tau_2$ ($a_2$)** | 2.1 (0.01) | 18.0 (0.32) |
| **$\tau_3$ ($a_3$)** | 18.0 (0.0007) | 81.4 (0.03) |
| **$\langle\tau\rangle$** | 2.6 | 29 |

**Table S1**: Best fit values calculated from the triexponential fitting of TRPL curves in Figure 2b.

# X-ray photoelectron spectroscopy (XPS)

## Methods

X-ray photoelectron spectroscopic (XPS) analyses were carried out in an UHV chamber with a base pressure lower than $10^{-9}$ mbar. The chamber was equipped with non-monochromatized Mg Kα radiation (hυ = 1253.6 eV) and a hemispherical electron/ion energy analyzer (VSW mounting a 16-channel detector). The operating parameters of the X-ray source were 12 kV and 12 mA and photoelectrons were collected normal to the sample surface, maintaining the analyzer angle between analyzer axis and X-ray source fixed at 54.5°. All the samples were drop casted on In foil and on a slab of Au on mica and XPS spectra acquired in a fixed analyzer transmission mode with pass energy of 44.0 eV. The spectra were analyzed by using the CasaXPS software. Linear or Shirley functions have been used to subtract the background. The de-convolution of the XPS spectra has been carried out applying a combination of Gaussian and Lorentzian functions (70:30). The binding energy scale was calibrated using the Au4$f_{7/2}$ peak or the In3$d_{5/2}$ peak respectively at 84 eV,[13] and 443.9 eV.[14]

## XPS detailed analysis

XPS technique provided a chemical characterization of the functionalized CdSe QDs and confirmed the formation of a S-Cd bond between the **1** molecules and the CdSe QDs, as briefly described in the main text. The analysis of the Cd3*d* and Se3*d* regions (see Figures S21 and S22) confirms nanoparticles composition in line with what previously reported in the literature:[15] Cd3*d* spectra features a signal at 404.8 eV and its relative spin orbit, while in Se3*d* region a signal at 53.4 eV is present. In Figure 2c (main text), the C1*s* region spectra acquired on bulk molecules, on pristine CdSe QDs and on QD-χ-C60 system are reported. While in the spectrum of pure $C_{60}$ only an intense signal at 285 eV arising from C=C species is found,[16] in the spectrum of **2** four components are needed to fit properly the experimental data, these components can be attributed to C-C (284.6 eV), C-N/C-O (286.5 eV), COOR (288.4 eV) and to a *shake-up* transition (290.3 eV).[17] The same components can be detected also in the spectrum of **1** with the exception of the one at lower binding energy which is

covered by the presence of an intense component at 285.1 eV attributable to the C $sp^2$ belonging to the fullerene moiety of the molecules.[16] As a further reference, the C1*s* spectrum of CdSe QDs has been also collected; here, a main component at 284.3 eV is observed attributable to aliphatic carbon atoms of the TOPO ligand, plus a minor component at 285.8 eV attributable to adventitious carbon. Finally, the C1*s* region acquired on the QD-χ-C60 sample features all the components observed in previous systems suggesting the coexistence of both **1** and TOPO molecules after the exchange reaction. A decrease of the component attributable to TOPO has been observed and the low value of the ratio between the latter and the components belonging to the **1** molecules clarify that the residual TOPO contamination is a minority part of the system. A further confirmation of the presence of residual TOPO after the exchange with **1** is given by the observation of a P2*p* contribution at 132.4 eV[18] in the XPS spectra acquired on samples after the exchange reaction (see Figure S20).

A piece of crucial information regarding the correct assembly of molecules on the surface of nanoparticles can be deduced also by the analysis of S2*p*/Se3*p* region (Figure 219). In the spectra of bulk molecules, a component attributable to thiol group at 163.5 eV and its relative spin orbit signal at higher binding energy (+1.18 eV) is observed.[19] In addition, a small component at ca. 167.1 eV is found and attributed to a partial oxidation of sulfur atoms of molecules.[20] In CdSe QDs sample a sulfur signal is not expected to be observed, while a Se3*p* signal (159.7 eV) and its relative spin orbit (+5.7 eV) are clearly visible.[15] In the QD-χ-C60 sample, due to the overlap of Se3*p* and S2*p* contribution, a change in the lineshape of the spectra is observed which is given by the contemporary presence of selenium and an additional components at 161.8 eV attributable to sulfur atoms bound to the surface of the QDs.[21] This corroborates the correct assembly of molecules on the surface. Furthermore, the spectra do not feature a contribution at 163.5 eV as well as at ca. 167 eV excluding the presence of both physisorbed and oxidated molecules.

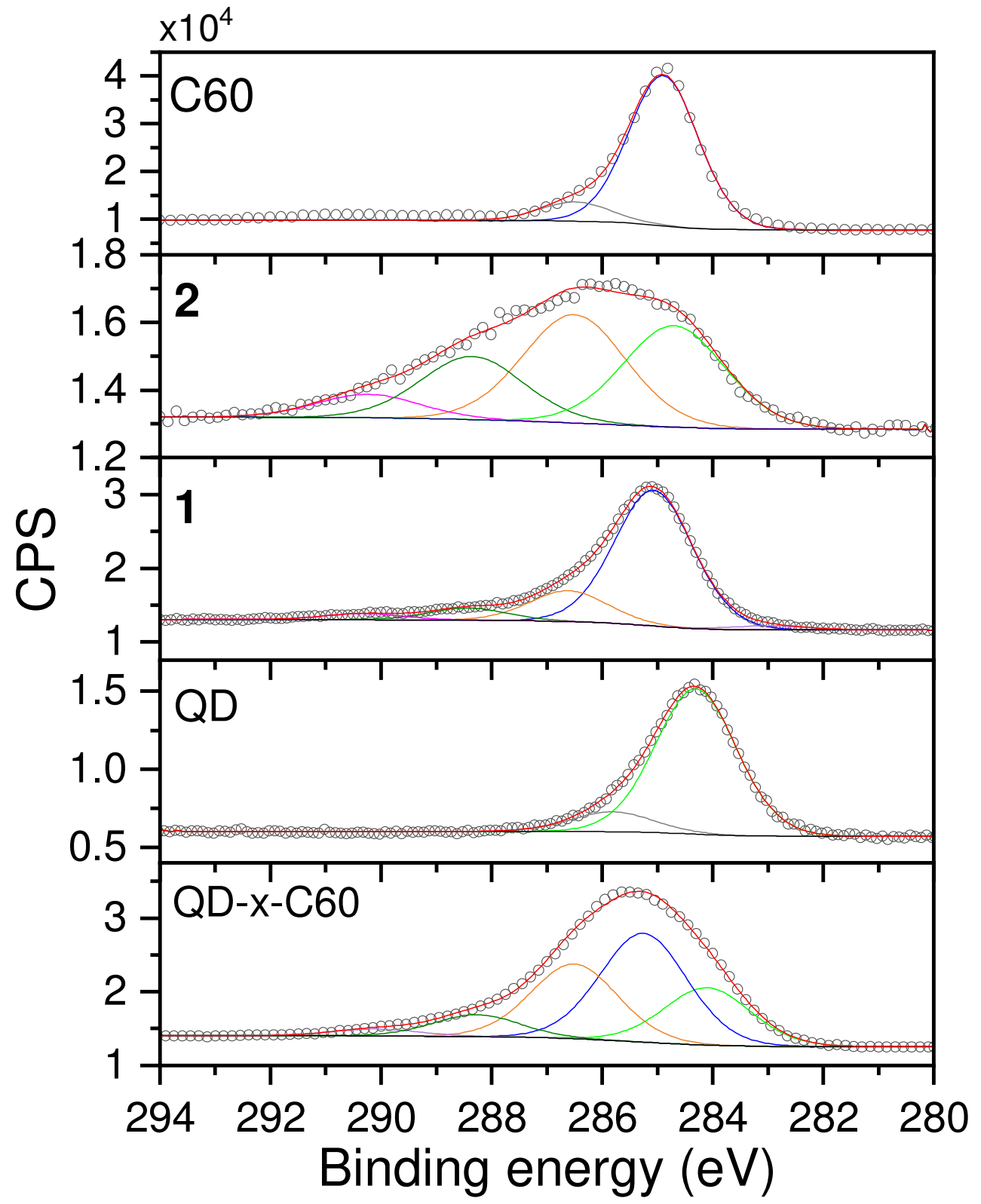


**Figure S18:** C1*s* photoemission lines for the $C_{60}$, the **2** and **1** ligands, the CdSe QDs and the QD-χ-$C_{60}$ system, as well as the single chemically shifted components from fit deconvolution.

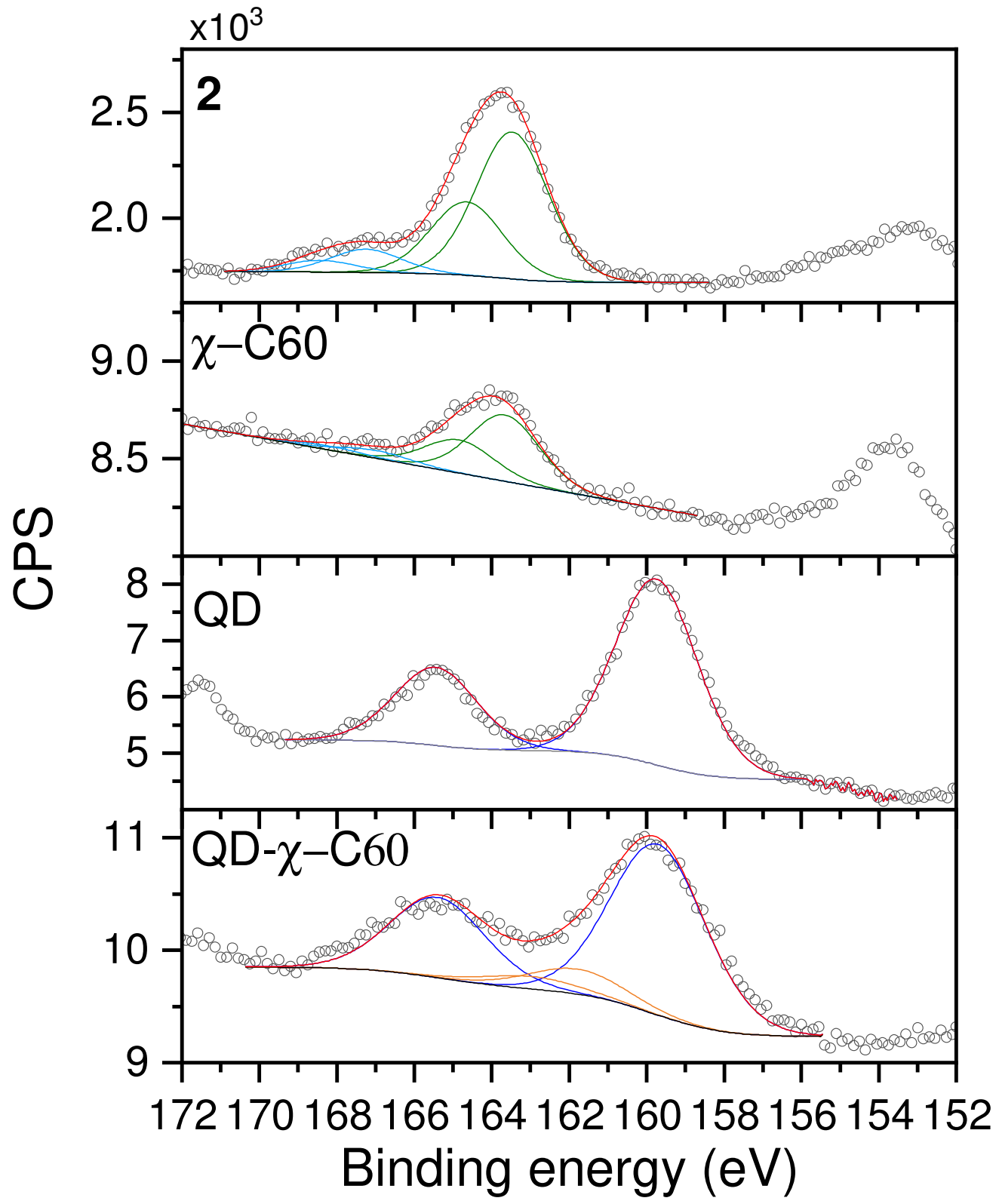


**Figure S19:** S2*p*/Se3*p* photoemission lines for the **2** and **1** ligands, the CdSe QDs and the QD-χ-$C_{60}$ system, as well as the single chemically shifted components from fit deconvolution.

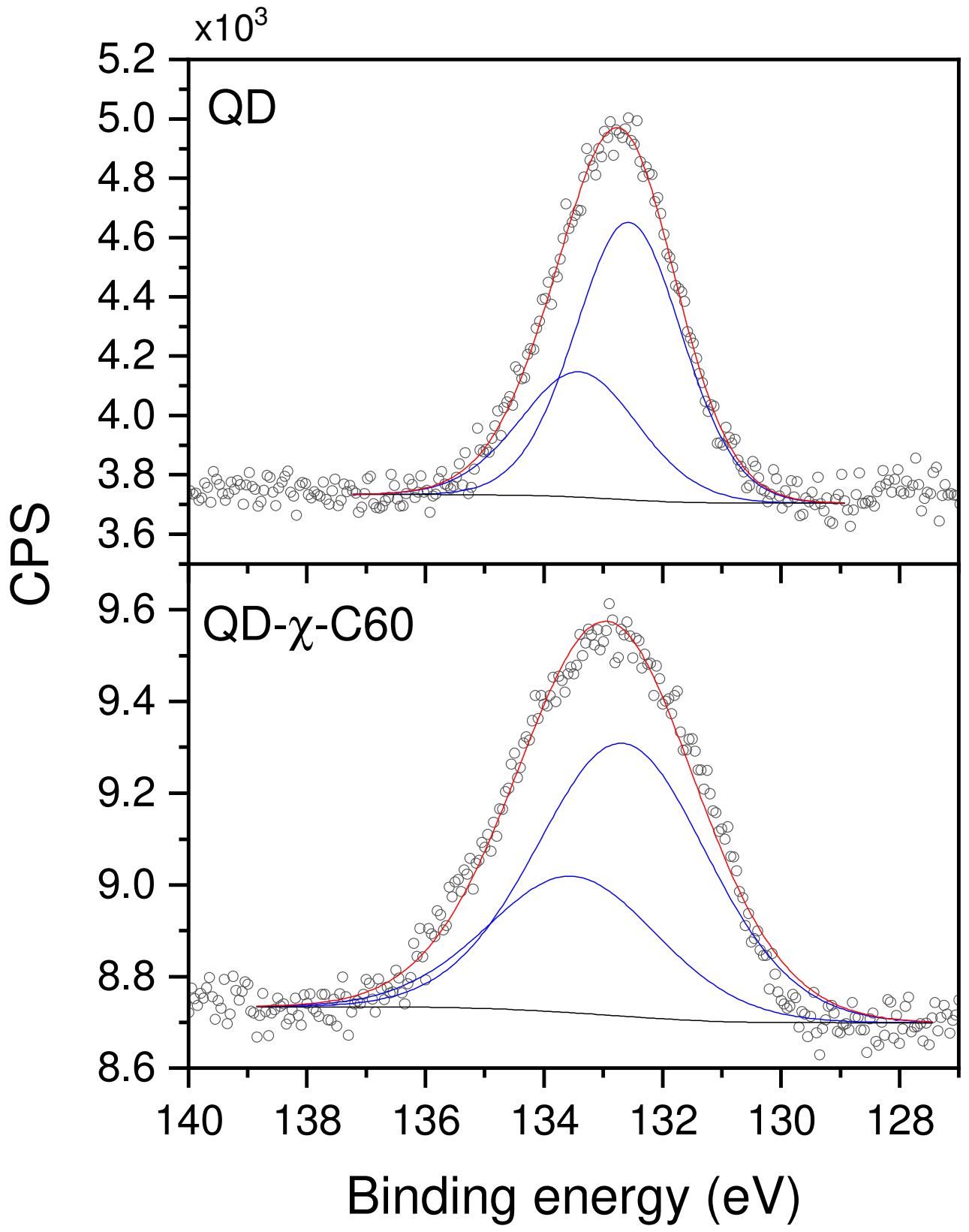


**Figure S20:** P2*p* photoemission lines for the CdSe QDs and the QD-χ-$C_{60}$ system, as well as the single chemically shifted components from fit deconvolution.

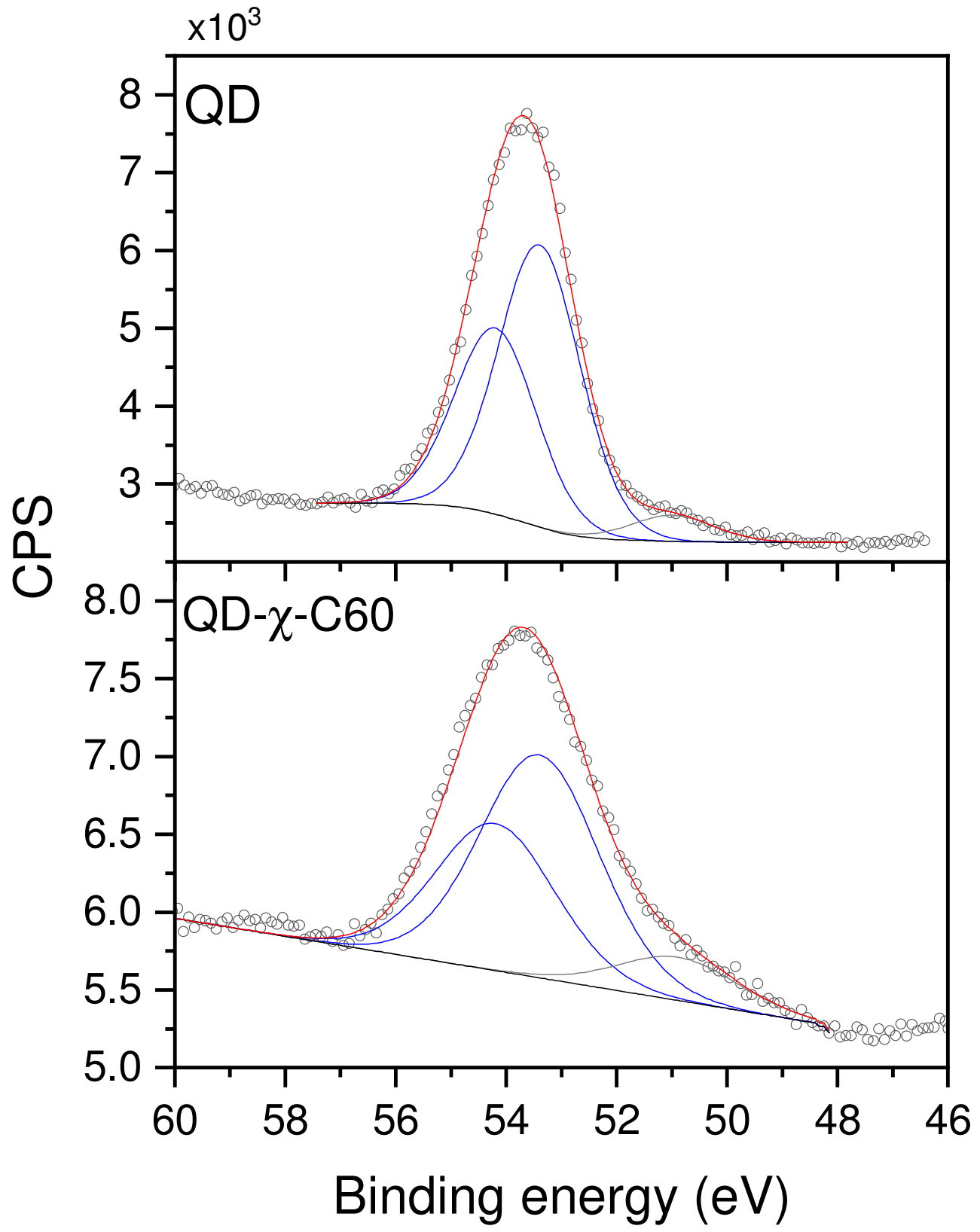


**Figure S21:** Se3*d* photoemission lines for the CdSe QDs and the QD-χ-$C_{60}$ system, as well as the single chemically shifted components from fit deconvolution.

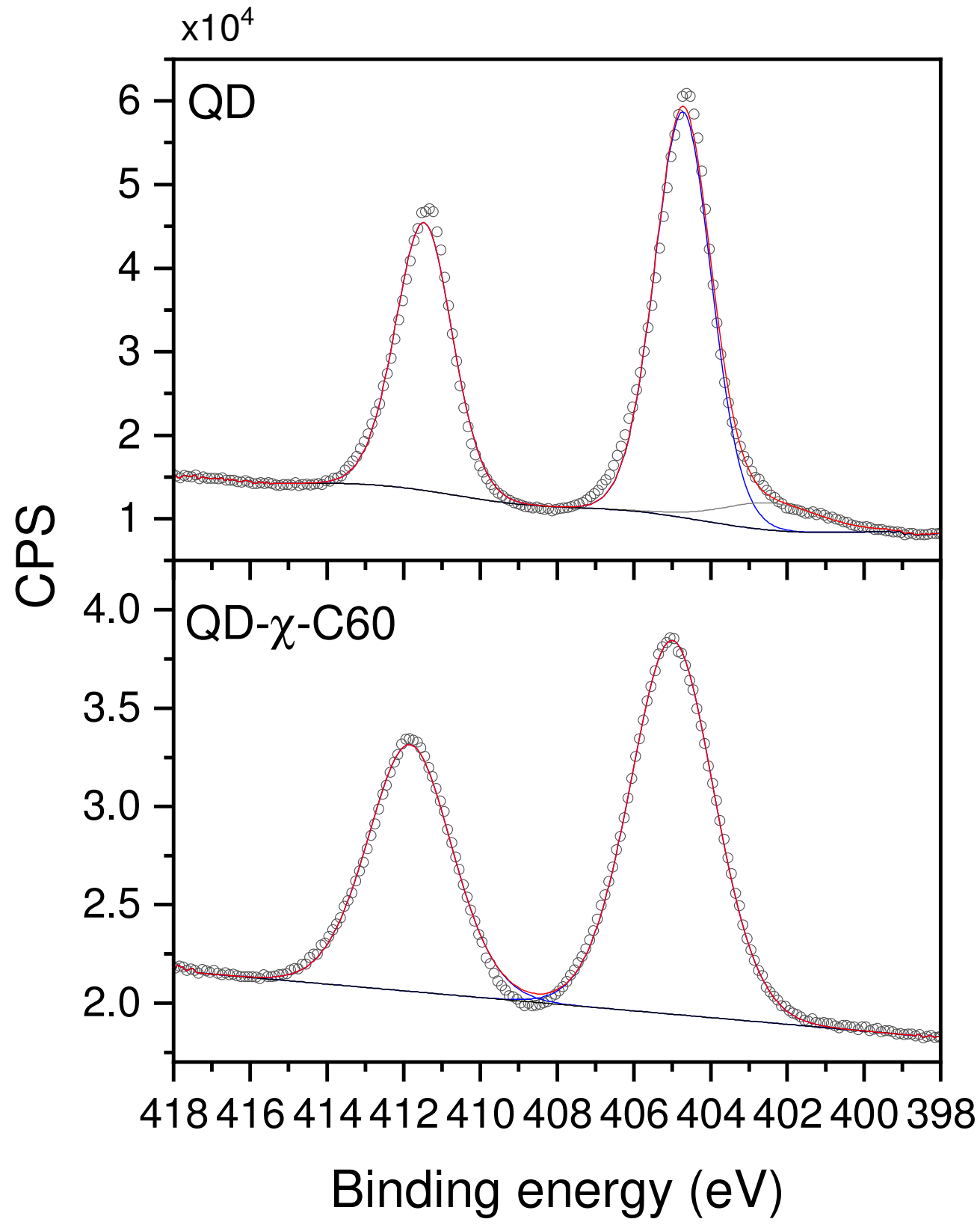


**Figure S22:** Cd3*d* photoemission lines for CdSe QDs and the QD-χ-$C_{60}$ system, as well as the single chemically shifted components from fit deconvolution.

# Time-resolved Electron Paramagnetic Resonance (trEPR)

## Methods

All trEPR spectra were recorded on a Bruker Elexsys E580 X-band spectrometer, equipped with a helium gas-flow cryostat for sample temperature control. The sample temperature was maintained with an Oxford Instruments CF9350 cryostat and controlled with an Oxford Instruments ITC503. Laser excitation at different wavelengths was provided by a Litron AURORA II opto-parametric oscillator (OPO) tunable laser (model number: A23-39-21, 21 Hz repetition rate, E/pulse ≈ 2 mJ, $\lambda$ = 410−700 nm, pulse duration = 7 ns). trEPR experiments were performed by direct detection with the transient recorder without lock-in amplification. The instrument response time was about 200 ns. The spectra were acquired with 2 mW microwave power and averaging 100 transient signals at each field position. The magnetic field was measured with a Bruker ER035M NMR gaussmeter.

The trEPR measurements were performed on the model system CdSe QD-χ-$C_{60}$ and the control sample, as a comparison, consisting of CdSe QD-χ to which 1 mM of PCBM was added. The concentration of CdSe QD was chosen in order to have an optical density of 1 in the 0.33 cm, i.e. 7.8 μM in 1,2,4-trichlorobenzene. The solutions were poured inside EPR quartz tubes that were sealed with Teflon under $N_2$ atmosphere.

From the data set obtained, the transient EPR spectrum at different time delays after the laser pulse was extracted. The reported trEPR spectra have been averaged over a time window of 0.2 μs. The acquired trEPR spectra have been simulated by using the core functions *pepper* and *esfit* of the open-source MATLAB toolbox EasySpin.[22] The parameters included in the best-fit simulations are the *g*-value (assumed isotropic), ZFS parameters (*D* and *E*), the triplet population sublevels ($p_x$, $p_y$, $p_z$) and the line broadening (assumed as Lorentzian). To carry out least-square fittings, a user-defined simulation function was developed, which allowed the fitting of the spin triplet sublevels populations (Figure S23 and Table S2). All the fits were carried out using a Nelder/Mead downhill simplex optimisation algorithm.

**trEPR best-fit simulations**

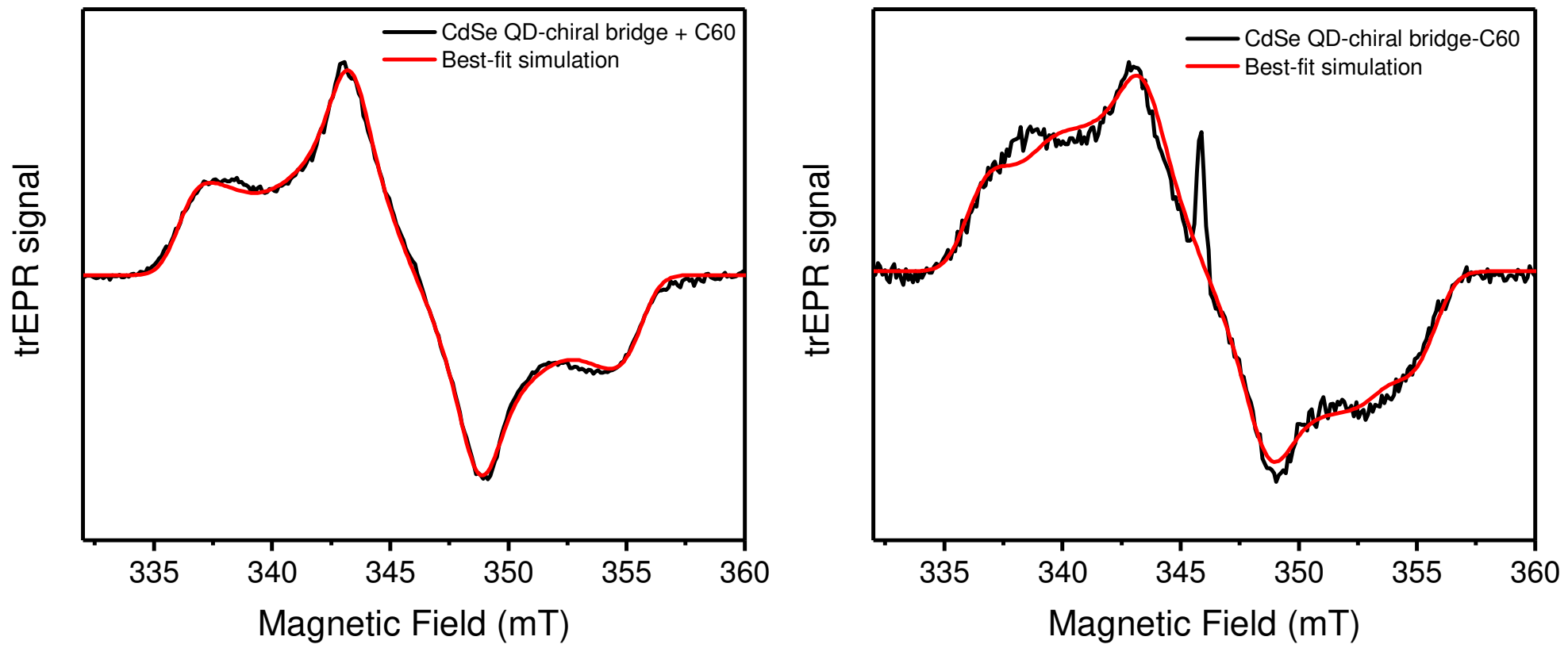


**Figure S23.** Best-fit spectral simulations of the trEPR spectra of CdSe QD-χ + $C_{60}$ (left) and CdSe QD-χ-$C_{60}$ (right) taken at 1 us after 450 nm laser pulse (7ns, 2mJ) acquired at 40 K. Only the triplet signal (not the narrow central radical signal) was considered in the simulation.

| | ***CdSe QD-χ +C₆₀*** | ***CdSe QD-χ-C₆₀*** |
|---|---|---|
| ***g*** | 2.000 | 2.000 |
| ***[D E]/MHz*** | [-276 40] | [-278 42] |
| ***LW/mT*** | 1.5 | 1.7 |
| ***[$p_x$ $p_y$ $p_z$]*** | [0.32 0.68 0.00] | [0.41 0.59 0.00] |

**Table S2.** A summary of the best-fit spectral simulations of the trEPR measurements reported in Figures S23. For each sample, the ZFS parameters of the triplet state, given in units of MHz, are reported. From the ZFS parameters, we assigned the triplet to the $C_{60}$ triplet in accordance with literature values.[23-24] Only one populating mechanism, i.e. SOC-promoted ISC, was considered in the simulations. Only Lorentzian broadening was considered to avoid an over-parametrization of the fitting; the linewidth is reported in units of mT.

# Theoretical modelling

## Simulation of incoherent time evolution

Spin relaxation, dephasing and charge recombination are introduced in our simulations through the super-operator $\boldsymbol{L}$ in the stochastic Liouville equation

$$\frac{\partial \rho}{\partial t} = -i[H, \rho] - \boldsymbol{L}\rho$$

By defining $\boldsymbol{L} = \boldsymbol{R} + \boldsymbol{K}$, we separate contributions from spin relaxation and dephasing (contained in the super-operator $R_{i,j,k,l}$), and charge recombination (in the super-operator $K_{i,j,k,l}$). The effect of the super-operators $R_{i,j,k,l}$ and $K_{i,j,k,l}$ is that of connecting the $\rho_{i,j}$ and the $\rho_{k,l}$ elements of the density matrix, where the indices $(i, j, k, l)$ run over the four eigenstates of the Hamiltonian. Given the large difference between the two g-factors, the eigenstates are practically factorized, hence $i, j, k, l \in \{\uparrow\uparrow, \uparrow\downarrow, \downarrow\uparrow, \downarrow\downarrow\}$ (here, z is the direction of the external field). This allowed us to introduce two different spin relaxation times $T_1^D$ and $T_1^A$, which are introduced in $R_{i,j,k,l}$ as:

$$R_{\uparrow\uparrow,\uparrow\uparrow,\uparrow\downarrow,\uparrow\downarrow} = \frac{1}{T_1^A\left(1 + e^{\frac{E_{\uparrow\uparrow} - E_{\uparrow\downarrow}}{k_B T}}\right)}, R_{\uparrow\downarrow,\uparrow\downarrow,\uparrow\uparrow,\uparrow\uparrow} = R_{\uparrow\uparrow,\uparrow\uparrow,\uparrow\downarrow,\uparrow\downarrow} e^{\frac{E_{\uparrow\uparrow} - E_{\uparrow\downarrow}}{k_B T}}$$

$$R_{\uparrow\uparrow,\uparrow\uparrow,\downarrow\uparrow,\downarrow\uparrow} = \frac{1}{T_1^D\left(1 + e^{\frac{E_{\uparrow\uparrow} - E_{\downarrow\uparrow}}{k_B T}}\right)}, R_{\downarrow\uparrow,\downarrow\uparrow,\uparrow\uparrow,\uparrow\uparrow} = R_{\uparrow\uparrow,\uparrow\uparrow,\downarrow\uparrow,\downarrow\uparrow} e^{\frac{E_{\uparrow\uparrow} - E_{\downarrow\uparrow}}{k_B T}}$$

$$R_{\uparrow\downarrow,\uparrow\downarrow,\downarrow\downarrow,\downarrow\downarrow} = \frac{1}{T_1^D\left(1 + e^{\frac{E_{\uparrow\downarrow} - E_{\downarrow\downarrow}}{k_B T}}\right)}, R_{\downarrow\downarrow,\downarrow\downarrow,\uparrow\downarrow,\uparrow\downarrow} = R_{\uparrow\downarrow,\uparrow\downarrow,\downarrow\downarrow,\downarrow\downarrow} e^{\frac{E_{\uparrow\downarrow} - E_{\downarrow\downarrow}}{k_B T}}$$

$$R_{\downarrow\uparrow,\downarrow\uparrow,\downarrow\downarrow,\downarrow\downarrow} = \frac{1}{T_1^A\left(1 + e^{\frac{E_{\downarrow\uparrow} - E_{\downarrow\downarrow}}{k_B T}}\right)}, R_{\downarrow\downarrow,\downarrow\downarrow,\downarrow\uparrow,\downarrow\uparrow} = R_{\downarrow\uparrow,\downarrow\uparrow,\downarrow\downarrow,\downarrow\downarrow} e^{\frac{E_{\downarrow\uparrow} - E_{\downarrow\downarrow}}{k_B T}}$$

We also introduced the relaxation terms between $\downarrow\uparrow$ and $\uparrow\downarrow$, which are defined based on a third relaxation time $T_{bc}$. For simplicity, in our simulations we assumed $T_{bc} = T_1^D$. The remaining diagonal elements of $\boldsymbol{R}$ are given by the stationary condition $\sum_j R_{jjii} = 0$.

Spin dephasing is also contained in the super-operator $\boldsymbol{R}$ and is defined as:

$$R_{i,j,i,j} = -\frac{1}{T_2} \,, i \neq j$$

Charge recombination was instead accounted for in $K_{i,j,k,l}$, defined on the singlet-triplet basis $i, j, k, l \in \{S, T_-, T_0, T_+\}$. The elements of $K_{i,j,k,l}$ are determined as:

$$K_{i,j,k,l} = K'_{i,k}\delta_{j,l} + K'_{j,l}\delta_{i,k} \,, K'_{i,j} = -\frac{1}{2}(k_S Q_S + k_T Q_T)_{i,j}$$

Where $k_S$ and $k_T$ are the charge recombination rates from the singlet state and from the triplet states, respectively; $Q_S$ and $Q_T$ are projectors onto the subspaces defined by the same states.

**CISS initial state**

The effect of CISS is included in our simulations as a "filter" applied on the initial state $\rho(0)$. This is achieved using the following projector, written on the $\{\uparrow\uparrow, \uparrow\downarrow, \downarrow\uparrow, \downarrow\downarrow\}$ local molecular basis:

$$P_{CISS} = \begin{pmatrix} 1 & 0 & 0 & 0 \\ 0 & 0 & 0 & 0 \\ 0 & 0 & 1 & 0 \\ 0 & 0 & 0 & 0 \end{pmatrix}$$

Which is used to generate the initial CISS state as $\rho_{CISS}(0) = P_{CISS}\, \rho(0)\, P_{CISS}$. Here, the transferred electron spin on the $C_{60}$ anion is forced to be in the $\uparrow$ orientation, whereas the spin on the QD retains its initial superposition.